\documentclass[useAMS,usenatbib]{mn2e}
\usepackage{amsmath,fleqn,graphicx,amssymb}
\usepackage{multirow}
\renewcommand{\[}{\begin{equation}}
\renewcommand{\]}{\end{equation}}
\def\p{\partial}

\def\ex#1{\left\langle#1\right\rangle}

\let\boldgrk=\gkvecten
\let\boldgrksc=\gkvecseven

\def\gkthing#1{{\mathchoice%
	{\hbox{{\boldgrk\char#1}}}
	{\hbox{{\boldgrk\char#1}}}
	{\hbox{{\boldgrksc\char#1}}}
	{\hbox{{\boldgrksc\char#1}}}}}

\def\vtheta{\gkthing{18}}

{\newif\ifnotend
\notendtrue
\def\veclist{ABCDEFGHIJKLMNOPQRSTUVWXYZabcdefghijklmnopqrstuvwxyz.}
\def\top#1#2.{#1}
\def\tail#1#2.{#2.}
\loop\expandafter\xdef\csname v\expandafter\top\veclist\endcsname%
{{\noexpand\bf\expandafter\top\veclist}}
\edef\veclist{\expandafter\tail\veclist}
\if\veclist.\notendfalse\fi\ifnotend\repeat}
\def\d{{\rm d}}

\def\cJ{{\cal J}}

\def\bolOm{\mbox{\boldmath$\Omega$}}
\def\vOmega{\bolOm}

\def\Gyr{\,\mathrm{Gyr}}
\def\Myr{\,\mathrm{Myr}}
\def\kpc{\,\mathrm{kpc}}

\def\kms{\,\mathrm{km\,s}^{-1}}

\def\msun{\,{\rm M}_\odot}

\def\pc{\,\mathrm{pc}}
\def\e{\mathrm{e}}

\def\TM{{\sc tm}}

\def\Omegap{\Omega_{\rm p}}

\title[Trapped orbits and solar-neighbourhood kinematics]
{Trapped orbits and solar-neighbourhood kinematics}

\author[James Binney]{
  James Binney$^1$\thanks{E-mail: binney@physics.ox.ac.uk}\\  
  $^1$Rudolf Peierls Centre for Theoretical Physics, Clarendon Laboratory,
  Parks Road,
  Oxford, OX1 3PU, UK
}

\begin{document}
\maketitle

\begin{abstract}
Torus mapping yields constants of motion for stars trapped at a resonance.
Each such constant of motion yields a system of contours in velocity space at
the Sun and neighbouring points. If Jeans' theorem applied to resonantly
trapped orbits, the density of stars in velocity space would be equal at all
intersections of any two contours. A quantitative measure of the violation of
this principal is defined and used to assess various pattern speeds for a
model of the bar recently fitted to observations of interstellar gas.
Trapping at corotation of a bar with pattern speed in the range
$33-36\Gyr^{-1}$ is favoured and trapping at the outer Lindblad resonance is
disfavoured. As one moves around the Sun the structure of velocity space
varies quite rapidly, both as regards the observed star density and the zones
of trapped orbits. The data seem consistent with trapping at corotation.
\end{abstract}


\section{Introduction} \label{sec:intro}

The release of kinematic data from Gaia in 2018 April allowed the study of
velocity space with unprecedented precision. Study of this space was opened
up by the first astrometric satellite, Hipparcos \citep{Perryman1997}. Hipparcos
did not provide line-of-sight velocities so \cite{WD98} mapped velocity
space using 5-dimensional data. The all-sky nature of the Hipparcos catalogue
enabled him to infer the probability density of stars in velocity space, but
individual stars could only be assigned locations in velocity space on the
release of data from the Geneva-Copenhagen Survey \citep[][hereafter
GCS]{GCS04,GCS09,Casagrande2011}, which measured, inter alia,
line-of-sight velocities $v_\parallel$.  Once the GCS data were available it
became clear that little of the structure in velocity space is due to
dissolving groups of coeval stars since prominent structures contained stars
widely spread in age and chemical composition \citep{Faea05}.

\cite{WD99:Bar,WD00:OLR} proposed that the `Hercules stream', the dominant feature
at low azimuthal velocity $V$, is caused by the outer Lindblad resonance
(OLR) of the Galactic bar.  More recently \cite{Perez2017}
have attributed the Hercules stream to the bar's corotation resonance, but
this association is controversial
\citep{MonariHerc2017,TrickFragkoudi2019,FragkoudiTrick2019}.
\cite{DSea04} investigated the possibility that much of the remaining
structure is caused by spiral arms, a possibility that was been examined
several times subsequently \citep[e.g.,][]{Se10,Hahn2011,PJM11:Hyades,PJM13:Hyades}. 

The last decade has seen growing awareness of the value of angle-action
coordinates for Galactic dynamics, in part due to the discovery of an
algorithm, the `St\"ackel Fudge' \citep{JJB12:Stackel}, for computing angles
and actions $(\vtheta,\vJ)$ from ordinary phase-space coordinates $(\vx,\vv)$
of particles moving in any realistic axisymmetric Galactic potential
$\Phi(R,z)$. Several papers
\citep[e.g.][]{BinneySchoenrich2018,JBH_Galah2019,TrickRix2019,Vasiliev2019,HuntBubBovy2019}
have used this algorithm to discuss distributions of angle-action coordinates
using data from Gaia's second data release \citep{GaiaKatz2018}.

The distribution of stars in three-dimensional action space has attracted the
most attention because, by the theorem of \cite{Jeans1916}, a relaxed Galaxy
would be fully described by this distribution. Moreover, each point in this
space is associated with three characteristic frequencies $\Omega_r$,
$\Omega_\phi$ and $\Omega_z$, and the planes on which these frequencies
satisfy a resonant condition $\vN\cdot\vOmega=m\Omegap$ are of interest. Here
$\vN$ is a vector with integer components, $m$ is another integer and
$\Omegap$ is the `pattern speed' of the Galactic bar or a system of spiral
arms.  Several attempts have been made, with mixed success, to identify features
in the computed density of stars in action space with such resonant planes
\citep{Monari2017,Monari2019,FragkoudiTrick2019,HuntBubBovy2019}.

Any system of angle-action coordinates is a valid system of canonical
coordinates for phase space and the distribution of stars in its action space
may be
of interest, but the system is most valuable if the Galaxy's Hamiltonian
$H(\vx,\vv)$ is a function $H(\vJ)$ of the actions alone. Hitherto stars have
been assigned locations in action space under the assumption that the
Galaxy's gravitational field is axisymmetric although the bar and spiral
structure ensure that it is not. Hence the coordinates employed do not make the
true Hamiltonian a function $H(\vJ)$. In this paper the mapping between
velocity space and action space is discussed using actions that make the
Hamiltonian of a realistic model of the Galaxy's barred gravitational field a
function $H(\vJ)$ of the actions only. 

Introduction of a rotating bar qualitatively changes the structure of action
space. In fact, through the phenomenon of resonant trapping, it splits action
space into disjoint pieces because there is a separate action space for each
family of trapped orbits. Previous papers in this series
\cite{Binney2016,Binney2018}  demonstrated that torus mapping
\citep[][and references therein]{JJBPJM16} in combination with resonant perturbation theory, enables
one to compute, with remarkable accuracy, the mapping
$(\vtheta,\vJ)\rightarrow(\vx,\vv)$ for any family of resonantly trapped
orbits. One can also compute the boundaries of the region in the action-space
of the underlying axisymmetric model that should be excised and replaced with
the action space of the trapped orbits. Outside this excised area
non-resonant perturbation theory can be used to modify actions
computed under the assumption of axisymmetry into actions that reflect the
presence of the bar.

Section~\ref{sec:background}  reviews the methodology by which maps
$(\vtheta,\vJ)\rightarrow(\vx,\vv)$ are computed for resonantly trapped
orbits. Section~\ref{sec:corot} discusses the structure in velocity space
of orbits trapped at corotation, while Section~\ref{sec:OLR} presents a similar
exercise for orbits trapped at the OLR. Section~\ref{sec:away} compares
predictions for trapping at corotation with Gaia data for velocity spaces at
points a kpc distant from the Sun. Section~\ref{sec:conclude} sums up.

\section{Background}\label{sec:background}

The work reported here relies on  torus mapping, which was introduced by
\cite{McGJJB90}, completed by \cite{JJBKu93} and brought to a high state of
development in the PhD thesis of Kaasalainen
\citep{Ka94,Ka95:closed,Ka95:chaotic,KaJJB94:MNRAS}. A publically available
implementation of the technique in C++ was released as the `Torus Mapper'
(hereafter \TM) by \cite{JJBPJM16}. The key idea is that orbital tori in
phase space (surfaces $\vJ=\hbox{constant}$) can be constructed by injecting
an analytically obtained `toy torus' into the Galaxy's phase space with a
canonical map that is adjusted to minimise the variance of the Galaxy's
Hamiltonian within the three-dimensional volume of the injected torus.  The
toy torus is obtained by solving the Hamilton-Jacobi equation by separation
of variables, typically under the assumption of a spherical potential. 

The injected torus is characterised by the numbers that define the toy torus
and the canonical transformation. Fewer than 100 numbers almost always
suffice. Once a torus has been constructed, not only can the position of a
star with the given actions be predicted at negligible cost arbitrarily far
in the past or the future, but one can determine whether a star will reach
any give location $\vx$, and, if it does, determine the velocities it will then
have, and the contribution it will make to the stellar density at $\vx$.
Consequently, a torus contains much more information than the time series one
obtains by integrating an orbit with a Runge-Kutta or similar algorithm.

Given a grid of injected tori in action space, intervening tori can be
quickly constructed by interpolation \citep{JJBPJM16}. This possibility plays
a key role in the construction of the tori of trapped orbits by resonant
perturbation theory. This theory uses a canonical transformation to
isolate a `slow angle' and its conjugate action.  One neglects the
dependence of the Hamiltonian on the remaining `fast' angles and
solves for the motion in the two-dimensional space of the slow angle and
action. Traditionally this is done using a pendulum equation. A companion
paper  \cite{Binney_negJ} explains why this approach breaks down in the case of Lindblad
resonances but can be fixed up within the context of \TM. 

Resonant
perturbation theory works exceptionally well in combination with torus
mapping for two reasons. First, torus mapping (which is not a perturbative
technique) enables one to construct an integrable Hamiltonian $H_0(\vJ)$ that
is much closer to the true Hamiltonian $H(\theta,\vJ)$ than one can come in
traditional applications of perturbation theory. Second, \TM\ allows one not
only to handle motion in the plane of the slow angle and action more
completely than hitherto, but also to include subsequently contributions from
the fast angles.

When an orbit becomes trapped, the actions conjugate to its two fast angles
remain (approximately) constants of motion, while the slow action can no
longer serve as a conserved quantity. Its place as an action is taken by the
action of libration $\cJ$ which is the amplitude of motion in the plane of
the slow angle and its action. The variable conjugate to $\cJ$ is the
angle of libration $\theta_\ell$, which evolves linearly in time with a new angular
frequency $\Omega_\ell$. 
There is an upper limit $\cJ_{\rm max}$ on the action of libration, and
$\Omega_\ell$ tends to zero as $\cJ$ tends to $\cJ_{\rm max}$. 

Details of how resonant perturbation theory works and is implemented in the
context of \TM\ can be found in \cite{Binney2016,Binney2018} and a companion
paper \cite{Binney_negJ}, which resolves a technical problem that emerged in the
course of this study. Here the focus is on comparing the map of local
velocity space that emerged from the RVS subset \citep{GaiaKatz2018} of the Gaia
DR2 release \citep{GaiaDR2general} with model maps computed by instances of the classes
{\tt resTorus\_c} and {\tt resTorus\_L} that were documented in
\cite{Binney2018} and \cite{Binney_negJ} and can be downloaded with \TM. The method {\tt
containsPoint} within both these classes returns the number of
velocities at which a torus reaches a given point $\vx$ together with the
values of $\vtheta$ associated with those visits and the Jacobian determinant
\[\label{eq:bigD}
D\equiv\biggl({\p(\vJ)\over\p(\vv)}\biggr)_\vx=\biggl({\p(\vx)\over\p(\vtheta)}\biggr)_\vJ
\]
that determines how much the orbit contributes to the stellar density at
$\vx$.

\subsection{Solar position and velocity}

Distances are taken from \cite{SchoenrichME2019}, who adopt $8.27\kpc$ as the
distance to the Galactic centre and $(U_0,V_0,W_0)=(11.1,250,7.47)\kms$ as
the Sun's Galactocentric velocity. According to this assumption, the Sun is moving
towards the Galactic centre, ahead of the local circular speed
$\Theta_0=239\kms$ and up out of
the plane.

\subsection{Units}

The units of mass, length and time are $1\msun$, $1\kpc$ and $1\Myr$. the
unit of velocity is then $\kpc\Myr^{-1}\simeq978\kms$. Actions have units of
$\kpc^2\Myr^{-1}$. Frequencies are plotted in units of
Gyr$^{-1}=0.978\kms\kpc^{-1}$.

\subsection{Galactic potential}

The Galactic potential is modelled as the sum of an axisymmetric part and a
modulation in azimuthal angle $\phi$ that carries no net mass. The
axisymmetric component is the potential fitted by \cite{PJM17} to a
variety of data tweaked to yield $R_0=8.27\kpc$ and $\Theta_0=239\kms$ by
increasing the central surface density and scale length of the stellar disc from
$8.96\times10^8$ to $9.2\times10^8\msun\kpc^{-2}$ and from $2.5$ to
$2.6\kpc$.

\def\rq{r_{\rm q}}
The bar's contribution to the potential is that fitted by \cite{SormaniIII}
to the flow of gas through the disc as mirrored in longitude-velocity plots
of HI and CO. The bar's density distribution is
\[
\rho_2(r,\vartheta,\phi)={A\over4\pi
G}\left({v_0\e\over\rq}\right)^2\e^{-2r/\rq}\sin^2\vartheta\cos(2\phi),
\]
where $A$ is the dimensionless strength of the bar, $\rq$ is its scale
length and $(r,\vartheta,\phi)$ are spherical polar coordinates. The bar's
potential is (Sormani private communication)
\begin{align}\label{eq:barS}
\Phi_2&(r,\vartheta,\phi)=\biggl({\e^{-2x}(2x^4+4x^3+6x^2+6x+3)-3\over20x^3}\cr
&-{x^2\over5}\hbox{E}_1(2x)\biggr)
A(v_0\e)^2\sin^2\vartheta\cos(2\phi),
\end{align}
where
$x\equiv r/\rq$ and $\hbox{E}_1(x)=\int_1^\infty\d t\,{\e^{-xt}/ t}$.

The azimuthal angle is measured relative to the bar's long axis, which
rotates at angular frequency $\Omegap$. \cite{SormaniIII} adopted
$v_0=220\kms$, and inferred from longitude velocity plots of HI
and CO that with  $A\ga0.4$ and $\rq\simeq1.7\kpc$. the results below are for
$A=0.4$ and $\rq=1.7\kpc$.

\section{Orbits trapped at corotation}\label{sec:corot}

At corotation $J_\phi$ is the slow action that is replaced by $\cJ$, and its
conjugate angle $\theta_\phi$ is replaced by the angle of libration
$\theta_\ell$. $J_r$ plays the role of the surviving fast action $J_3'$.

\begin{figure}
\centering
\includegraphics[width=.8\hsize]{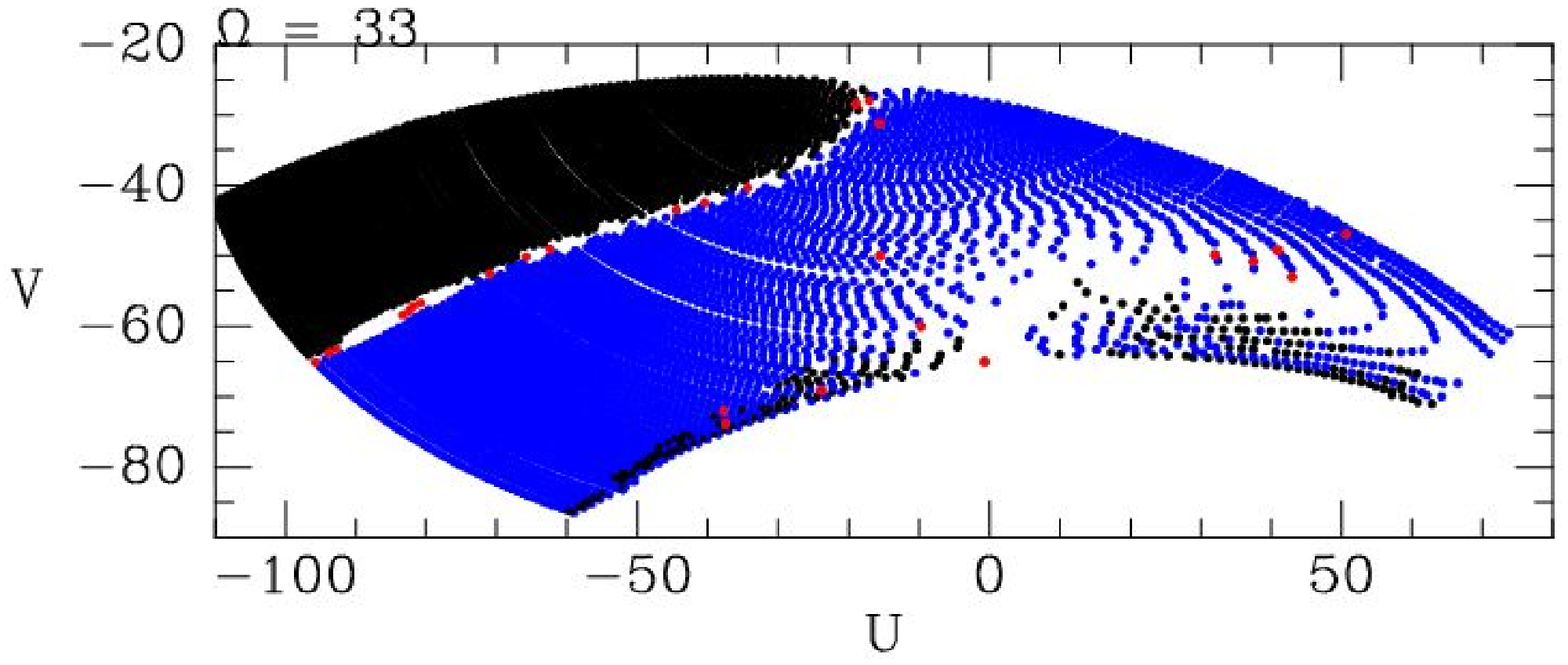}
\includegraphics[width=.9\hsize]{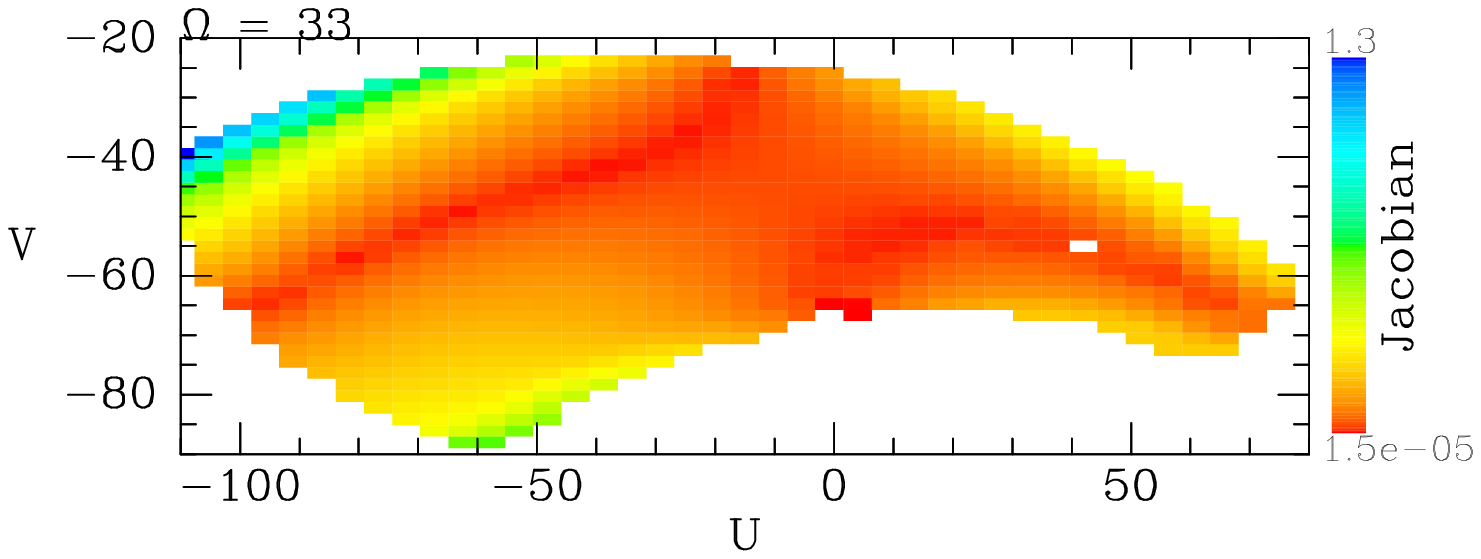}
\includegraphics[width=.9\hsize]{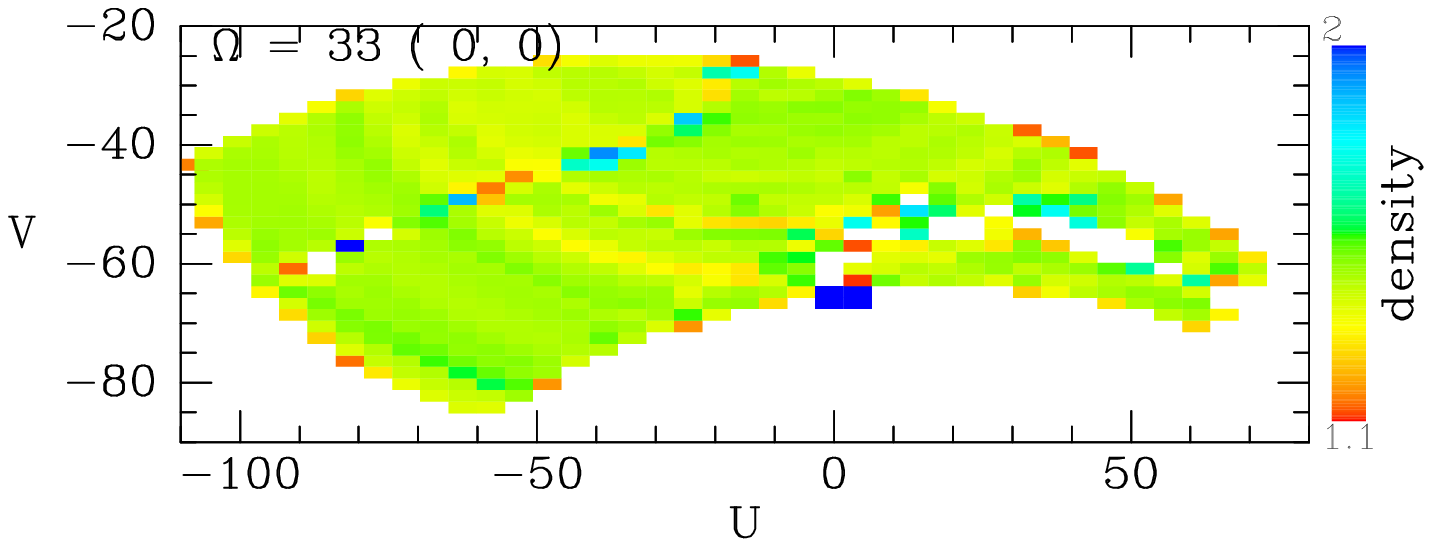}
\includegraphics[width=.8\hsize]{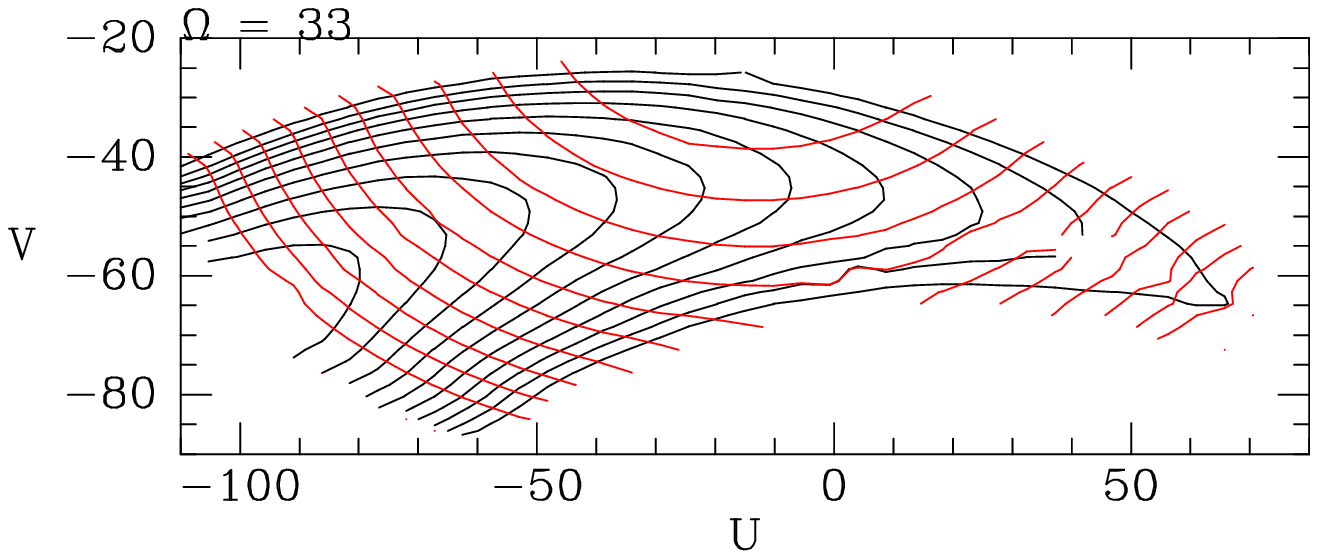}
\caption{The location in local velocity space of orbits trapped at corotation
by a bar with pattern speed $33\Gyr^{-1}\simeq33\kms\kpc^{-1}$. The top
panel shows the velocities at which trapped orbits pass through the Sun. Most
orbits reach us with two velocities (with $v_z>0$); one velocity is plotted
in black and the other in blue. The second panel down shows the value of the
Jacobian (\ref{eq:bigD}). The third panel down shows (on a logarithmic scale)
the (essentially
uniform) density in velocity
space generated by the way we have sampled action space. The bottom panel
shows contours of constant $J_r$ in red and constant $\cJ$ in black. The
range of libration actions is $(0.053,0.16)\kpc^2\Myr^{-1}$.
}\label{fig:corot33}
\end{figure}

Fig.~\ref{fig:corot33} shows the location in local velocity space of orbits
trapped by the Sormani et al.\ bar if
$\Omegap=33\Gyr^{-1}$. These orbits were obtained by
sampling the  action-space plane $J_z=0.0025\kpc^2\Myr^{-1}$ on  a
non-uniform rectangular
grid. The chosen value of $J_z$ causes stars to make excursions $\sim300\pc$
above and below the plane, but plays no essential role; any similar choice
would yield equivalent results. We restrict attention to instants when a star
passes the Sun moving upwards.

The $J_r$ coordinates of the grid points  were uniformly spaced in $\surd J_r$, a
choice which yields velocities which lie on approximate ellipses in the
$(U,V)$ plane that have near uniformly spaced semi-axis lengths. 
The $\cJ$ coordinate of the  $n$th grid point was
\[
\cJ_n=\cJ_{\rm min}+(n/N)^2[\cJ_{\rm max}-\cJ_{\rm min}].
\]
Here $N$ is the total number of $\cJ$ values to be sampled and $\cJ_{\rm
min}$ is the least libration action required for an orbit of a given value of
the fast action $J_3'=J_r$ to reach the Sun. $\cJ_{\rm max}$ is the
largest possible value of $\cJ$ at the given $J_3'$ value. 

The overwhelming majority of the 3611 orbits (tori) plotted in
Fig.~\ref{fig:corot33} visit the Sun at two different velocities. In the top
panel one velocity is marked by a blue dot and the other is
marked by a black dot.  14 orbits visit at a single velocity and 13
visit with three velocities. These velocities are marked by red dots. The
remaining 191 orbits visit with four velocities and two of these velocities
are plotted in blue and two in black.

\subsection{Density of stars in velocity space}

The second panel down shows how the Jacobian $D$ (eqn.\ref{eq:bigD}) varies
over the plane. It ranges over several orders of magnitude. It is smallest
where the density of dots in the top panel is lowest, namely where the blocks
of blue and black dots meet, and in the middle right of the wing structure.
To understand this connection we write
\[
\biggl({\d n\over\d^3\vv}\biggr)_\vx
={\d n\over\d^3\vJ}\,\biggl({\p(\vJ)\over\p(\vv)}\biggr)_{\vx}
=f(\vJ)\biggl({\p(\vx)\over\p(\vtheta)}\biggr)_{\vJ}=Df(\vJ).
\]
Hence when $D$ is small, the density of dots in velocity space will also be
small.
The third panel down in Fig.~\ref{fig:corot33} shows the density of stars in
velocity computed by adding to the grid of cells, via the cloud-in-cell
algorithm, an amount $1/D$ times the
area $\d\cJ\d J_r$ of the cell in the $(\cJ,J_r)$ plane that
is represented by that orbit. Thus $f(\vJ)$ is being set to a constant and
the plotted density is 
\[
f(\vJ)={1\over D}{\d n\over\d^3\vv},
\]
which should be constant. The third panel shows that to a good approximation it is

The attainment of a constant velocity-space density through cancellation of
the large variations in $D$ and in the dot density is striking. The physical
significance of these variations is that where the dots are sparse in the top
panel, $1/D=\p(\vtheta)/\p(\vx)$ is large because a given torus (almost)
reaches the Sun for a wide range of angle variables. Hence parts of velocity
space coloured red in the second panel of Fig.~\ref{fig:corot33} are
populated by large contributions from a small number of tori.

When comparing observations with distributions in velocity space like those
of Fig.~\ref{fig:corot33}, one has to be mindful that the observed star
density is not $\d n/\d^3\vv$ but $f[\vJ(\vx,\vv)]$ because it is $\d
n/\d^3\vv$ weighted by the real-space density of stars $\d n/\d^3\vx$, which
tends to infinity when $\d n/\d^3\vv$ tends to zero.

\subsection{Contours of constant action}

The bottom panel of Fig.~\ref{fig:corot33} shows contours of constant $J_r$
(red) and constant $\cJ$ (black). $J_r$ increases from top to bottom, and
$\cJ$ increases from left to right along the centre of the wing-shaped
region. The two sets of contours become tangent to one another along the
curve on which the blocks of black and blue dots meet in the top panel. This
parallelism indicates that the matrix
${\p v_i/\p J_j}$
that connects the $\vv$ and $\vJ$ coordinate systems for velocity space has
become degenerate. This does not imply any failure of the $(\vtheta,\vJ)$
system; it just signals that one needs to specify the value of an angle
variable in addition to the value of $J_r$ (or $\cJ$) to specify
unambiguously a velocity. As $\cJ$ decreases, the black contours shrink onto
a point that lies near $(-100,-70)\kms$. This is the velocity of the orbit
that has the minimum libration action ($\sim0.053$) required to reach the
Sun: $\cJ=0$ for the closed resonant orbit, which does not pass through the
Sun.

The red and black contours are also tangent in the lower right of the wing,
where the dots are very sparse in the top panel.

\begin{figure}
\centering \includegraphics[width=.9\hsize]{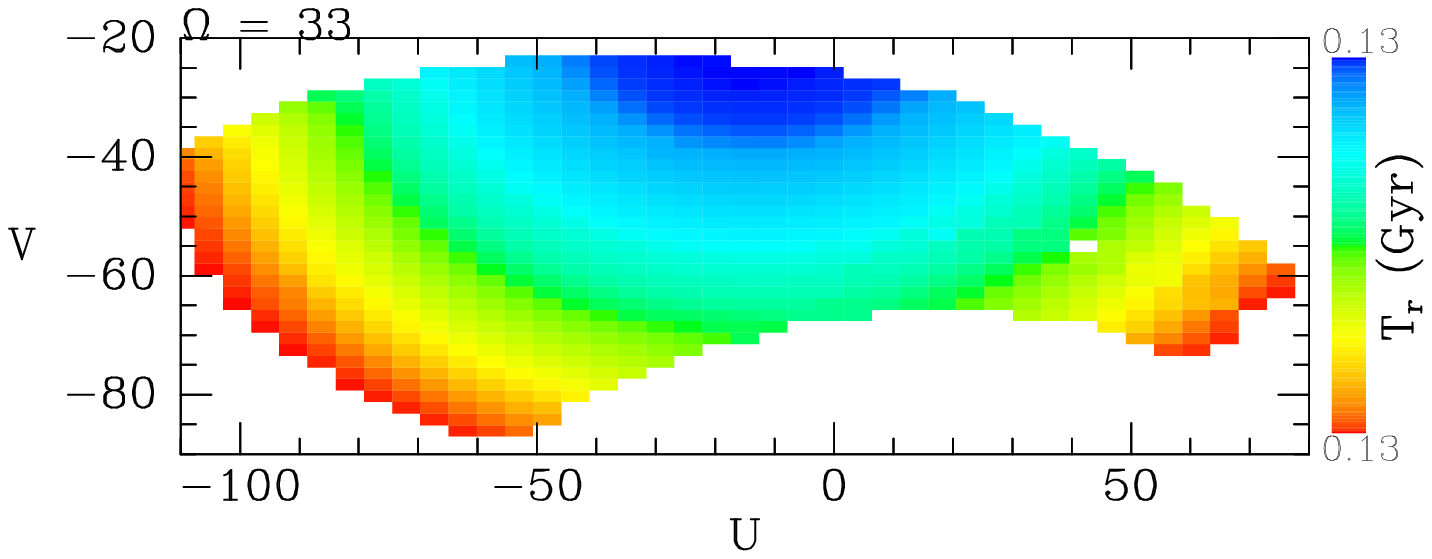}
\includegraphics[width=.9\hsize]{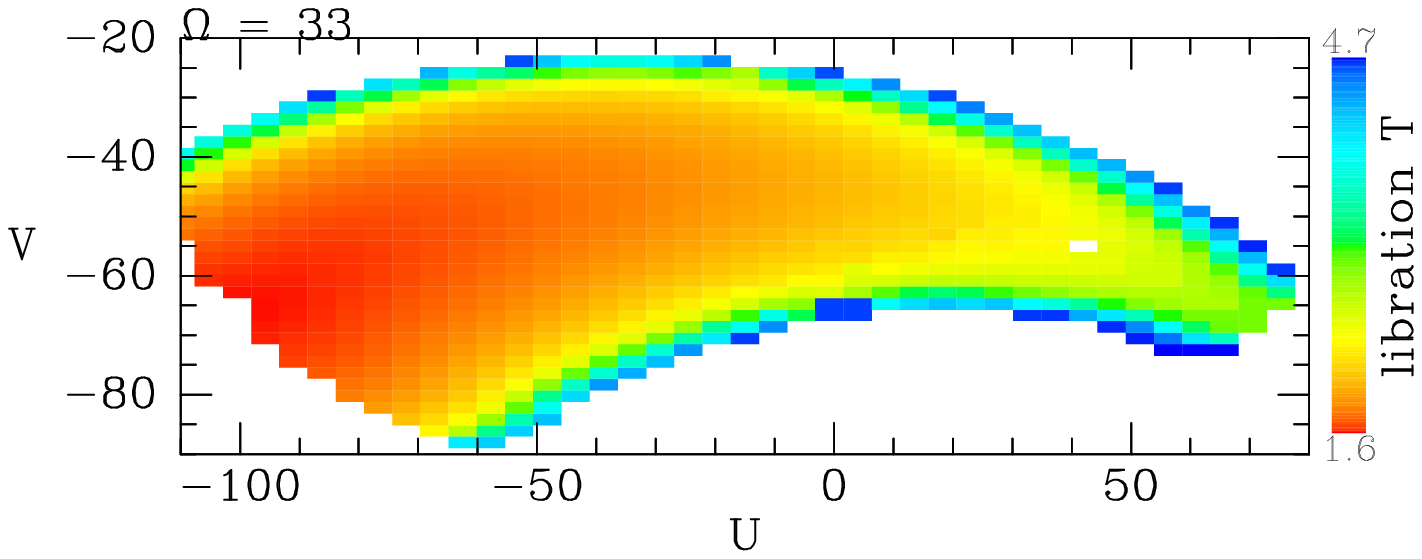}
 \caption{The radial period (upper) and libration period (lower) of orbits
trapped at corotation as a function of solar-neighbourhood velocity. Periods
are given in Gyr.}\label{fig:corot33T}
\end{figure}

Fig.~\ref{fig:corot33T} shows how the radial and libration periods
$2\pi/\Omega_r$ and $2\pi/\Omega_\ell$ vary within velocity space. The radial
periods lie within the very narrow range $130$ to $132\Myr$ while the libration
periods, which increase with libration amplitude, are all longer by more
than an order of magnitude. The shortest periods are $\sim1.6\Gyr$ and the
longest plotted exceed $4.6\Gyr$. Strictly,  the periods go to infinity at the
edge of the trapping zone, but the increase with $\cJ$ is finally very
rapid and  orbits with very long libration periods are probably
unphysical.

\begin{figure}
\includegraphics[width=\hsize]{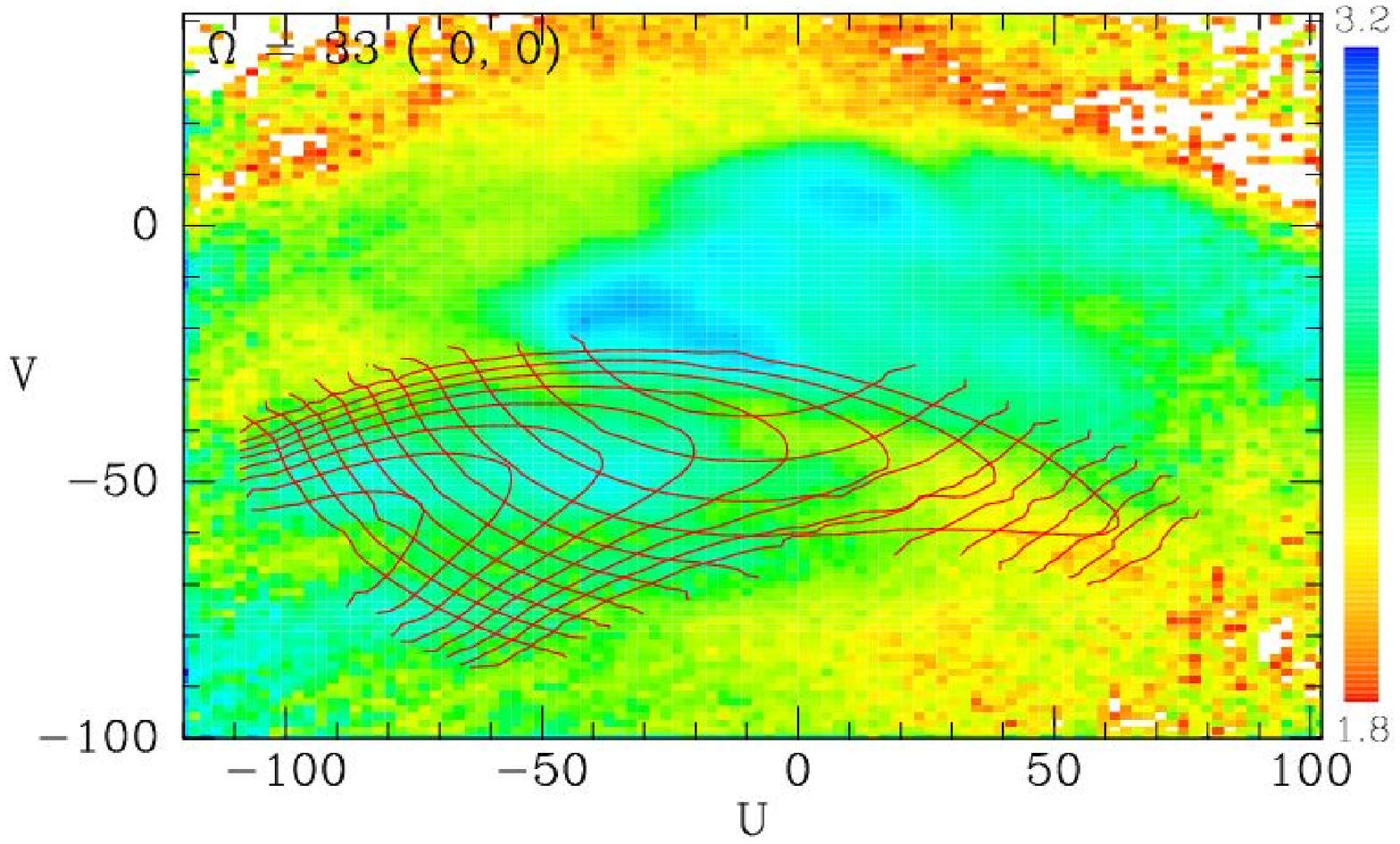}
\includegraphics[width=\hsize]{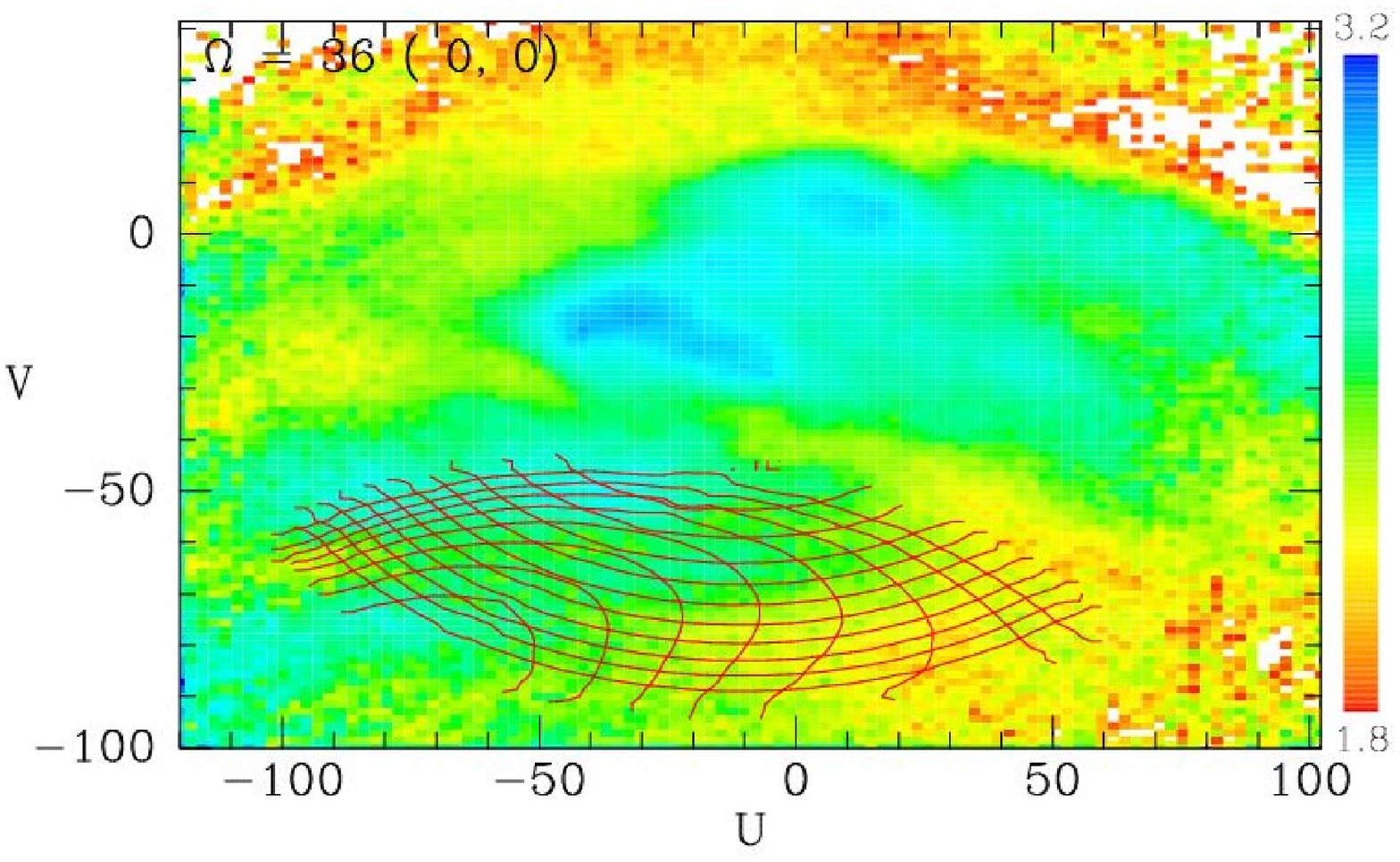}
\includegraphics[width=\hsize]{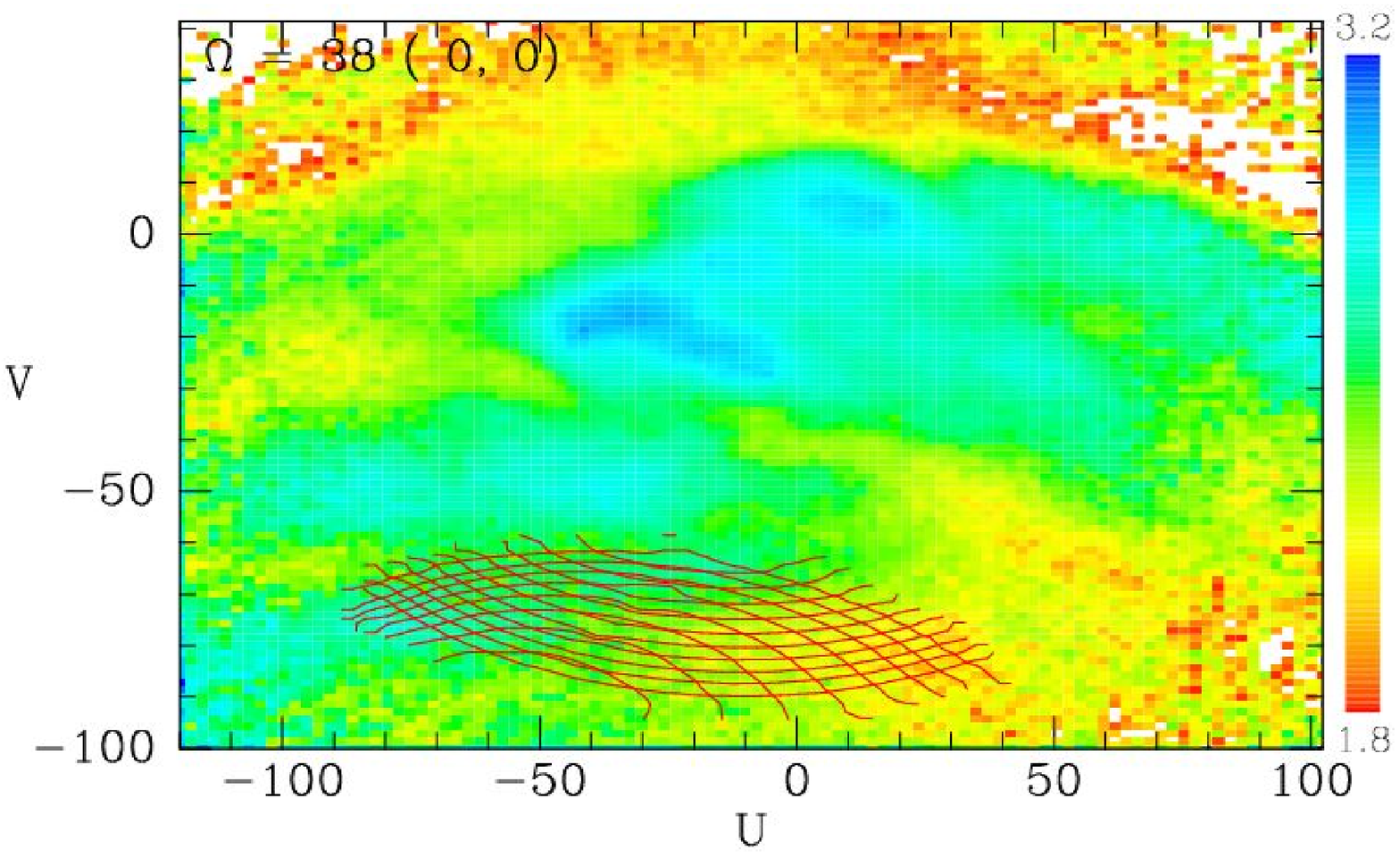}
 \caption{The contours of constant actions for orbits trapped by a bar with
pattern speeds $33$ (top) $36$ (middle) and $38\Gyr^{-1}$ superposed on the
density of stars in velocity space obtained by selecting stars that lie
within $0.3\kpc$ and have line-of-sight velocities from the RVS. The quantity
plotted is the logarithm to base 10 of the ratio of the measured density to
the prediction of an analytic DF.}\label{fig:Gaia33corot}
\end{figure}

\subsection{Comparison with Gaia data}

The top panel of Fig.~\ref{fig:Gaia33corot} superposes the contours of
constant action from the bottom panel of Fig.~\ref{fig:corot33} onto a
representation of the density of stars in the Gaia catalogue that lie within
$0.3\kpc$ of the Sun and have line-of-sight velocities in the Gaia DR2
catalogue.  This representation is designed to bring out local fluctuations
in the star density that are easily masked by the large contrast in the stellar
density between $(U,V)\simeq(0,0)$ and the edge. Specifically, the plotted
quantity is the logarithm to base 10 of ratio of the measured star density to
the density given by an
analytic DF for the discs (thin and thick) and the stellar halo.

In Fig.~\ref{fig:Gaia33corot} the region
covered by the contours is more extensive than the part of the plane that
lies below the depression that runs across the figure from centre left to
lower right. However, if one trimmed the contours back to include only orbits
with libration periods shorter than $\sim2\Gyr$, they would fit the region
below the depression rather nicely. The middle panel of
Fig.~\ref{fig:Gaia33corot} superposes contours for $\Omegap=36\Gyr^{-1}$.
The contours now seem unconnected to the data. The bottom panel, which shows
contours for $\Omegap=38\Gyr^{-1}$ reveals an alternative fit to the data:
now the contours provide a reasonable fit to a depression in the observed
star density that lies below the one that might be fitted by
$\Omegap=33\Gyr^{-1}$.

\subsection{Implication of Jeans theorem}
\begin{figure}
\centerline{\includegraphics[width=.8\hsize]{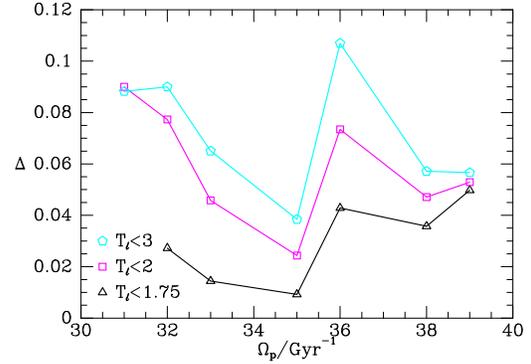}}
\caption{The statistic (\ref{eq:defDelta}) for different bar pattern speeds and
maximum libration periods $T_\ell$. Each symbol corresponds to a maximum libration
period as marked in Gyr.}\label{fig:sets}
\end{figure}

In Fig.~\ref{fig:Gaia33corot}, most contours of constant $\cJ$ cut several
contours of constant $J_r$ twice. Hence we can identify sets of velocities
that correspond to identical actions.  If Jeans' theorem were to apply, the
DF, and therefore the density of stars in the $UV$ plane, would take the same
value at every point in a given set of points. We can easily test this
conjecture since our tori visit just such matched points, and at these points
we should find equal densities of Gaia stars.  Fig.~\ref{fig:sets} shows the
statistic
\[\label{eq:defDelta}
\Delta\equiv{\ex{{\overline{n^2}-\overline{n}^2-\overline{n}\over\overline{n}^2}}}.
\]
 Here $n$ is the number of Gaia stars in the $UV$ bin picked out by a
visiting torus and an overline indicates a mean over the points visited by a
single torus while the angle brackets indicate a mean over all relevant tori.
The first two terms in the fraction's numerator give the variance in the
number of stars in the bins visited by a single torus and the third term
corrects for the contribution of Poisson noise. If Jeans's theorem were
perfectly observed, the numerator would average to zero. 
Three points are shown for each value of $\Omegap$, being the points obtained by
including only tori with libration periods shorter that $1.75$, $2$ and
$3\Gyr$ on the grounds that Jeans theorem is most likely to apply to the tori
with the shortest libration periods. The figure shows that in general longer
maximum libration periods yield larger values of $\Delta$ as expected, but the scatter
this induces does not obscure systematic variation of $\Delta$ with
$\Omegap$. For every  maximum period $\Delta$ takes its minimum value at
$\Omegap=35\Gyr^{-1}$, and the smallest upper limit on $T_\ell$ favours values
in the range $32-36\Gyr^{-1}$.

\section{Orbits trapped at OLR}\label{sec:OLR}

At OLR the slow angle is $\theta_1'=\theta_r+2\theta_\phi$. The value of this linear
combination of $\theta_r$ and $\theta_\phi$ is set by the action of libration
$\theta_\ell$.  The fast action
is $J_3'=J_\phi-2J_r$ and it is complemented by the action of libration
$\cJ$. 

\begin{figure}
\centering
\includegraphics[width=.8\hsize]{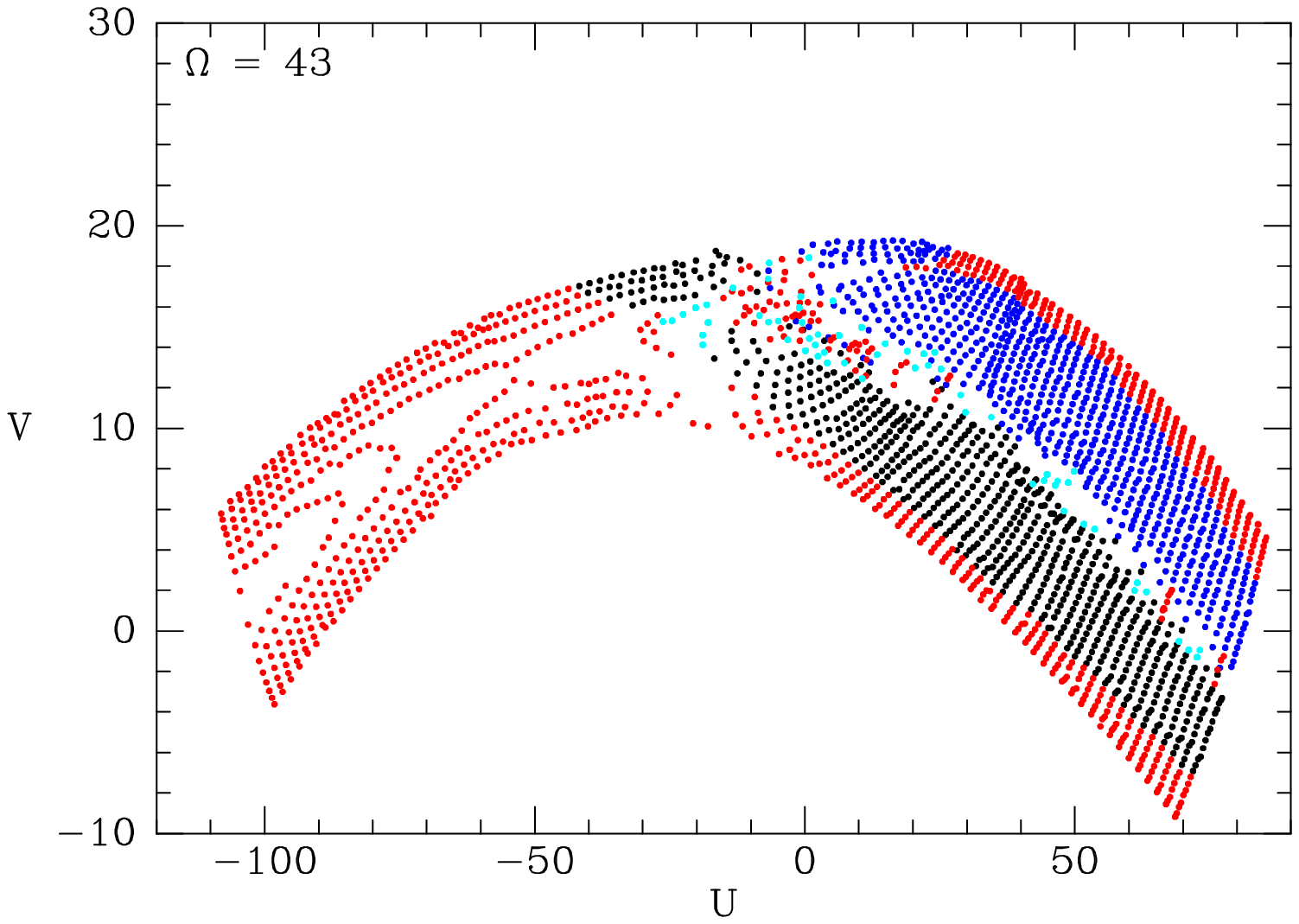}
\includegraphics[width=.8\hsize]{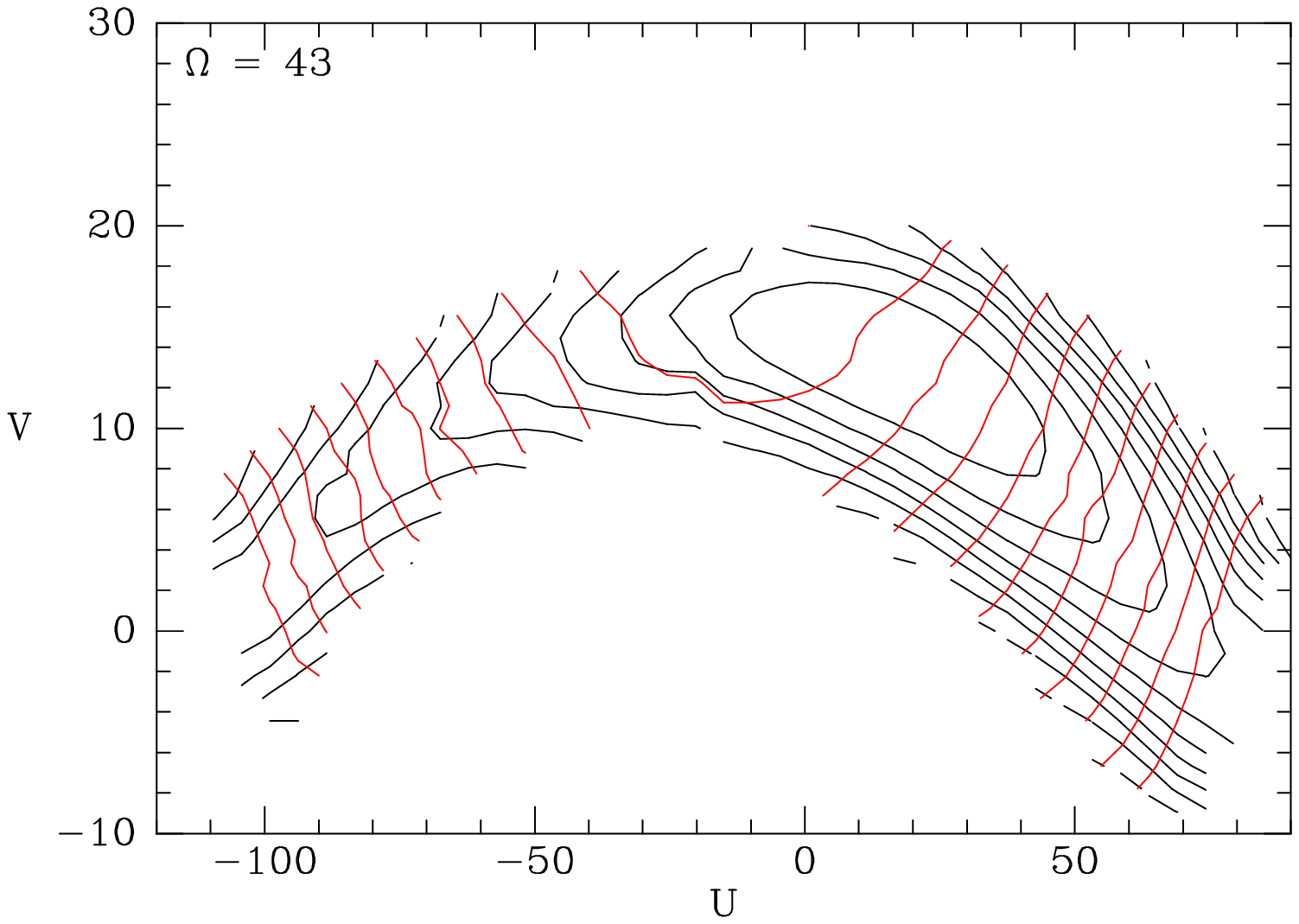}
\includegraphics[width=.9\hsize]{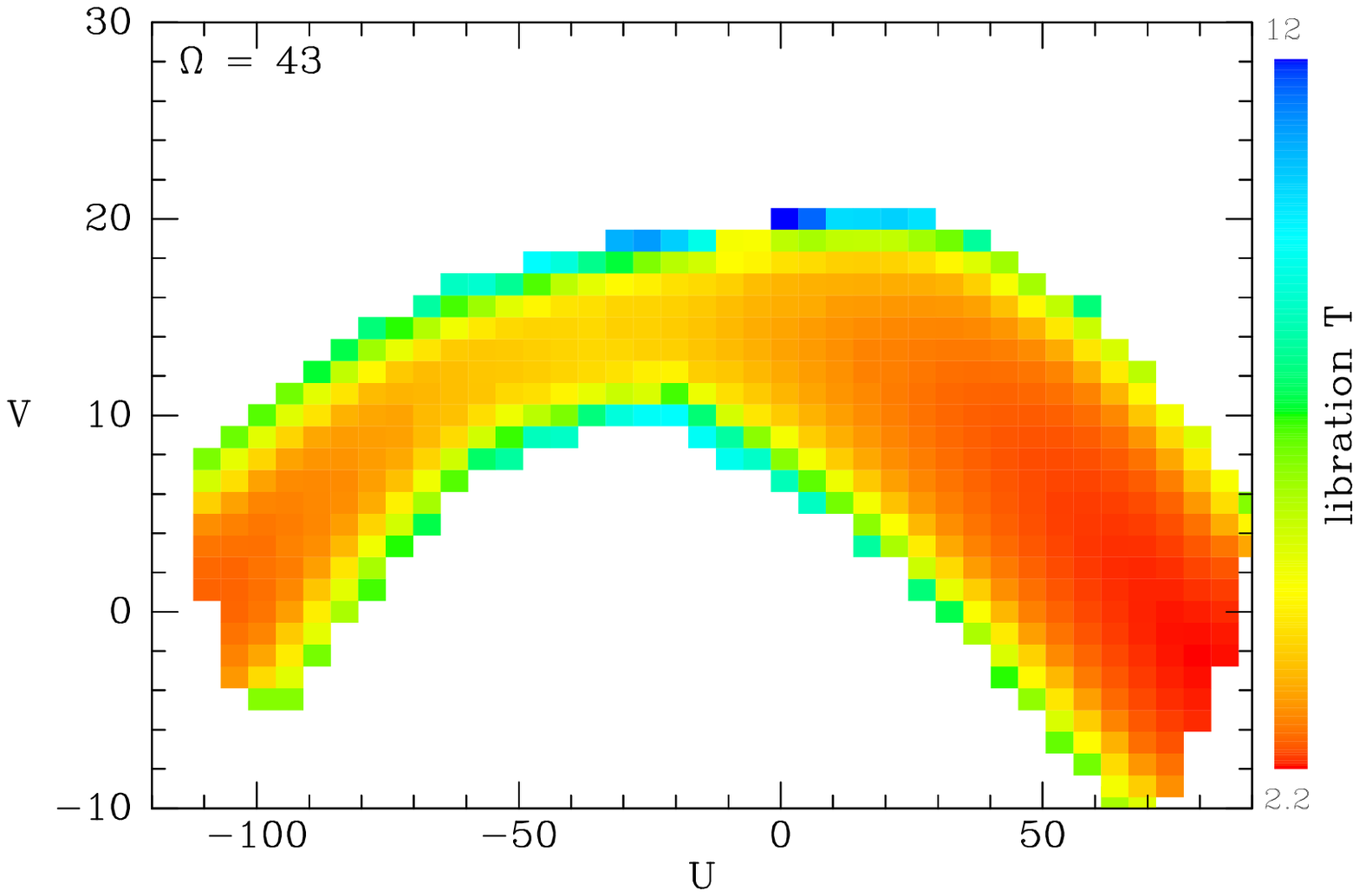}
 \caption{The locations in local velocity space of orbits trapped at OLR of a
bar with pattern speed $43\Gyr^{-1}$. The top panel shows the velocities at
which trapped orbits pass through the Sun. Most orbits reach us with either
two (black/blue points) or four (red points) velocities (with $v_z>0$). Cyan
points mark single or triple visits.  In the second panel down red curves are
contours of the conserved action $J_3'=J_r-2 J_\phi$ while contours of
constant libration action $\cJ$ are black. The values of $\cJ$ increase with
the length of the contour, while the values of $J'_3$ increase roughly with
$|U|$.  The bottom panel shows the means in each cell of the libration
period. Note that, unlike in Fig.~\ref{fig:corot33} the vertical and
horizontal scales differ.}\label{fig:OLR38}
\end{figure}

Fig.~\ref{fig:OLR38} shows the footprint in the $UV$ plane of orbits that are
trapped at the OLR with $\Omegap=43\Gyr^{-1}$. As in the case of corotation,
the great majority of trapped orbits visit the Sun (with $v_z>0$) at either
two or four points in the $UV$ plane: in this case 682/1000 visit twice
(blue/black dots) and 220 visit four times (red dots). The middle panel of
Fig.~\ref{fig:OLR38} explains the increased popularity of four visits:
several red contours (of constant $J_\phi-2J_r$) cut  black contours (of
constant $\cJ$) four times each, whereas in the bottom panel of
Fig.~\ref{fig:corot33} each red contour cuts a black contour at most twice.
We saw above that orbits correspond to intersections of contours, so four
intersections implies four visit. Stars that visit twice have either the
smallest $J_r$ or the smallest $\cJ$ consistent with reaching the Sun.

The bottom panel of Fig.~\ref{fig:OLR38} shows that all  orbits contributing
to Fig.~\ref{fig:OLR38} have
long libration periods: periods start at $2.2\Gyr$.

\begin{figure}
\centering
\includegraphics[width=\hsize]{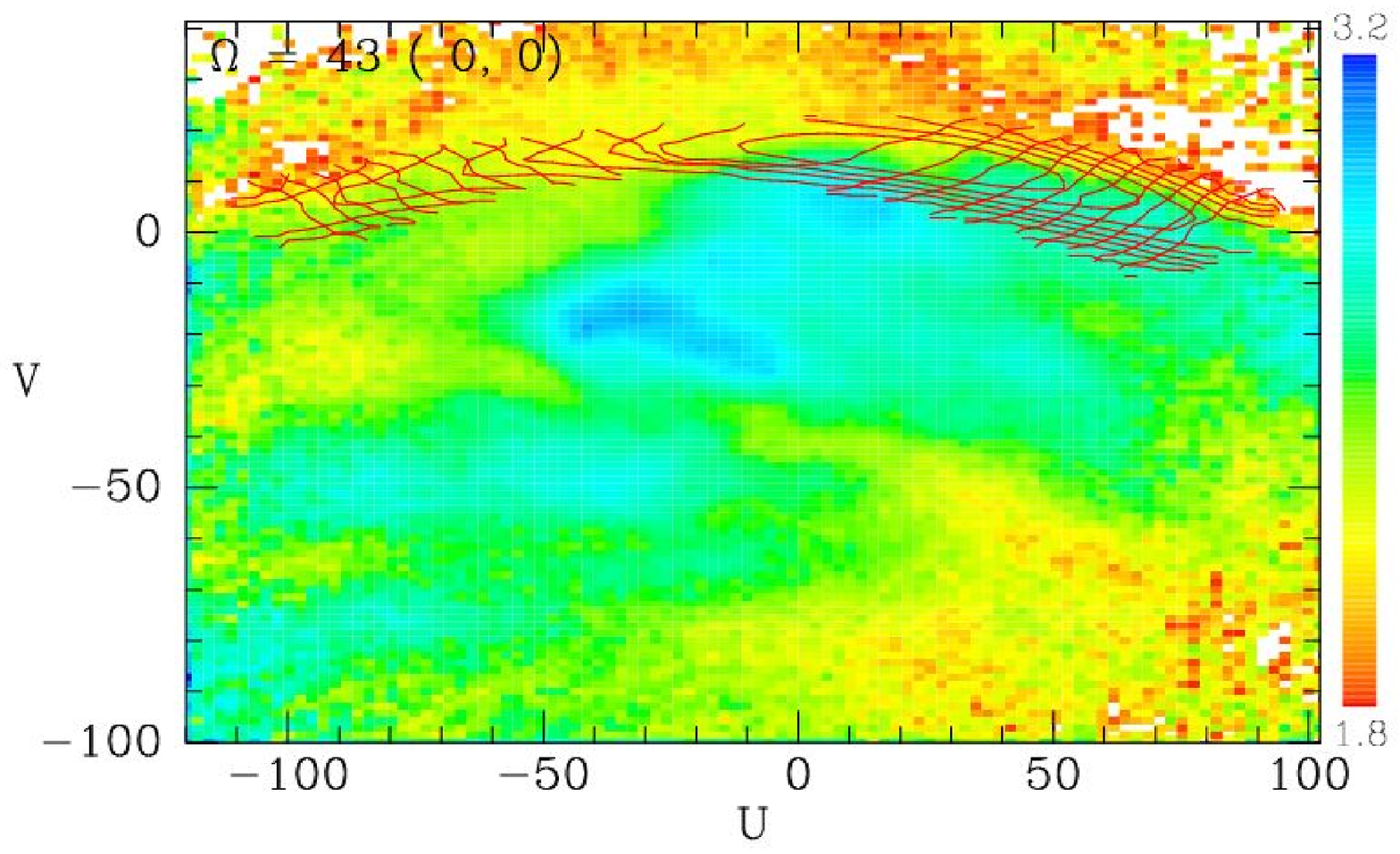}
\includegraphics[width=\hsize]{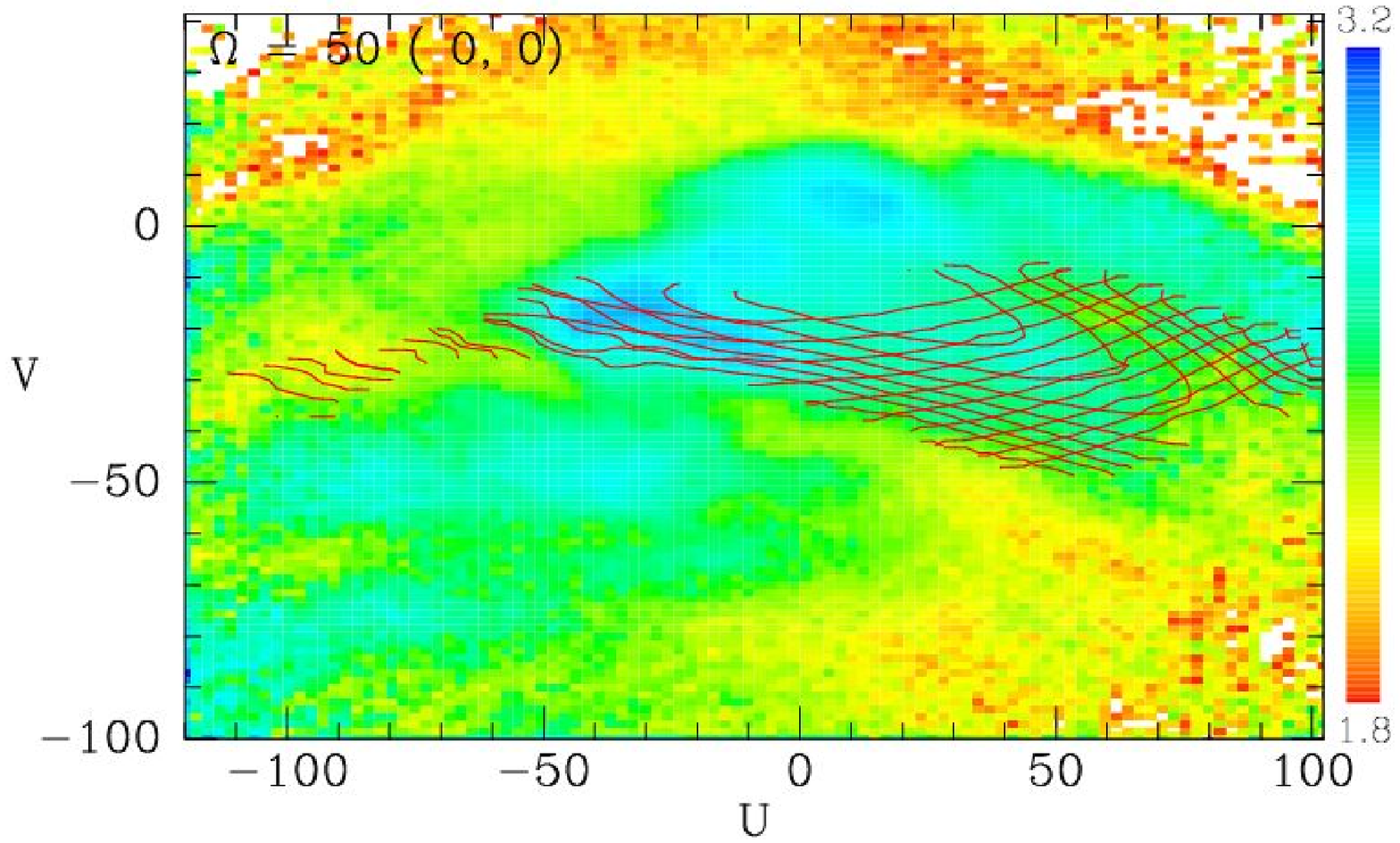}
\includegraphics[width=\hsize]{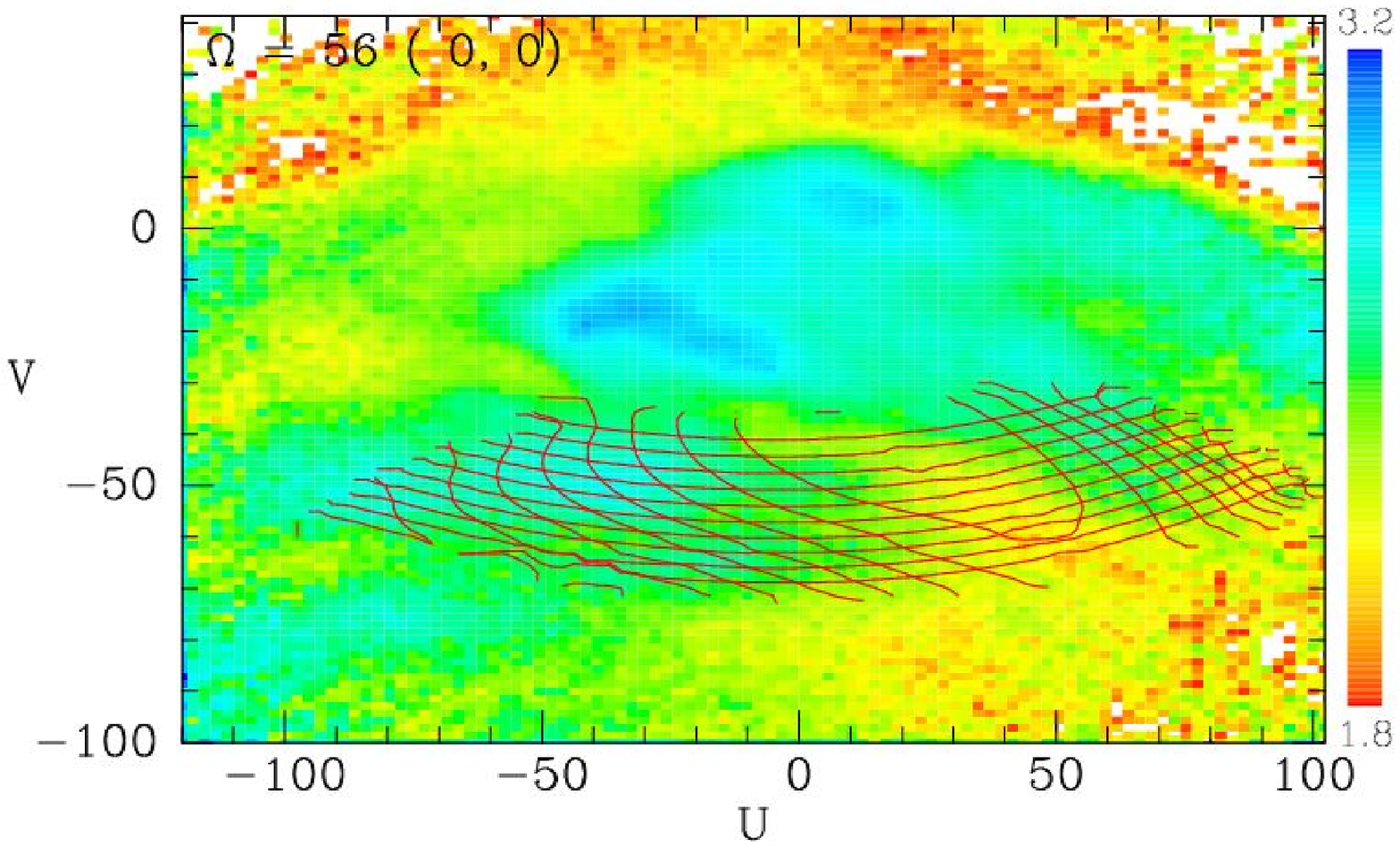}
\caption{The  density of RVS stars in local velocity space as in
Fig.~\ref{fig:Gaia33corot} together with
contours of constant actions for orbits trapped at the OLR of bars with
pattern speeds (from top to bottom) $\Omegap=38$, $50$ and
$56\Gyr^{-1}$.}\label{fig:DR2OLR}
\end{figure}

\begin{figure}
\centerline{\includegraphics[width=.8\hsize]{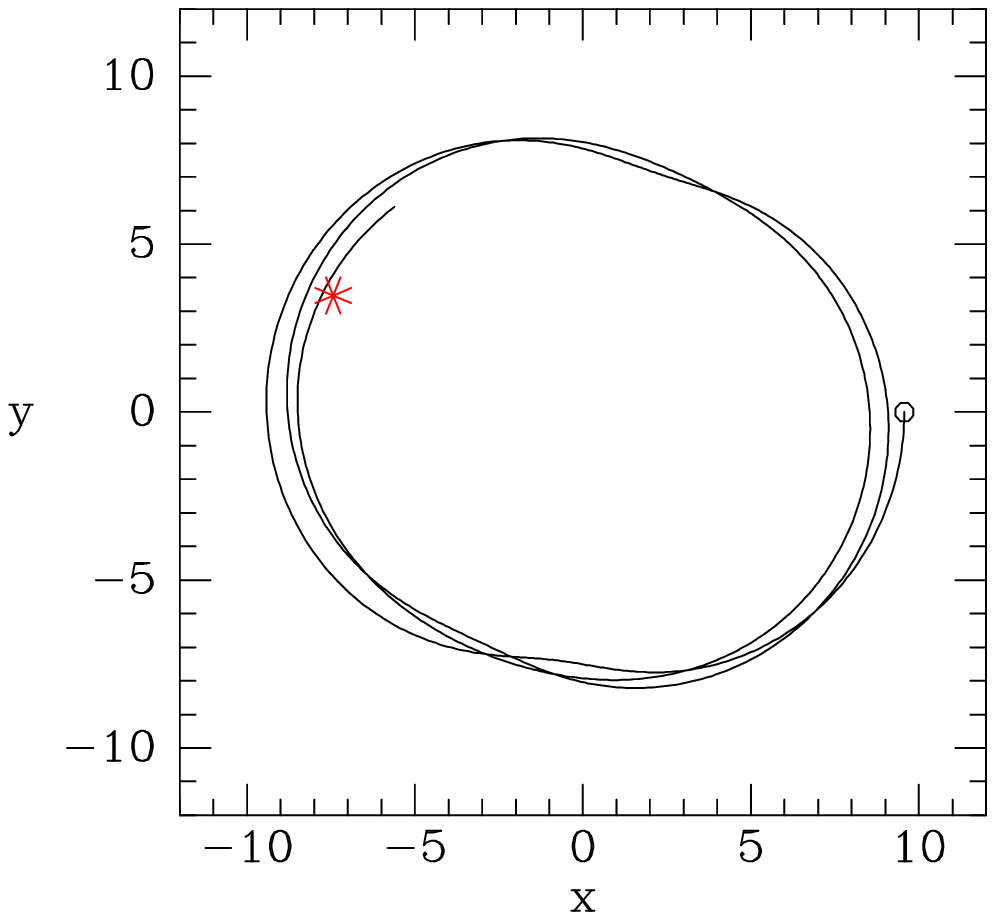}}
\caption{A section of an orbit trapped at OLR in a bar with $\Omegap=50\Gyr^{-1}$.
The circle marks the launch point and the star marks the Sun's
location.}\label{fig:OLR50orb}
\end{figure}

Fig.~\ref{fig:DR2OLR} overlays contours like those plotted in the middle
panel of Fig.~\ref{fig:OLR38} on the representation of the density of
stars in the Gaia DR2 RVS sample used in Fig.~\ref{fig:Gaia33corot}. For $\Omegap\la37\Gyr^{-1}$ the region
occupied by trapped orbits lies at values of $V$ that are too large to be of
interest. Consequently, if trapping at corotation is relevant, as
Fig.~\ref{fig:Gaia33corot} suggests, trapping at OLR will be unimportant. In
Fig.~\ref{fig:DR2OLR} the region occupied by trapped orbits moves down as
$\Omegap$ increases.  The upper boundaries of the regions of trapping in the
lower two panels of Fig.~\ref{fig:DR2OLR} are occupied by orbits with the
largest values of $J_3'$ and thus the smallest values of $J_r$. That is, the
top edges of these panels lie on contours of constant $J_3'$.

The middle panel of Fig.~\ref{fig:DR2OLR} shows that when
$\Omegap=50\Gyr^{-1}$ they reach the Sun with small negative values of $V$,
and for the most part positive values of $U$.  Fig.~\ref{fig:OLR50orb}
explains the bias to positive $U$ (movement towards the Galactic Centre) by
showing part of an orbit with a small value of $\cJ$ together with the
location of the Sun. In an inertial frame the bar and the disc rotate
counter-clockwise, but pictured here in the bar's frame the orbit circulates
clockwise, and as it passes the Sun, it is moving from apocentre at the bar's
major axis towards pericentre near the minor axis.  Hence orbits with small
$\cJ$ visit us with $U>0$.

Of the 2139 orbits computed for $\Omegap=50\Gyr^{-1}$, 1795 visit the Sun
twice with $v_z>0$ and just 246 visit three or four times. Thus increasing
$\Omegap$ has made four visits less popular. The stars that still visit four
times make up the tail of contours in the middle panel of
Fig.~\ref{fig:DR2OLR} that extends leftwards of $U=-60\kms$. The tail is made
up of stars with large libration actions that also visit on the upper and
lower edges of the band of contours at $U>0$ (cf.\ the top panel of Fig~\ref{fig:OLR38}).

The bottom panel of Fig.~\ref{fig:DR2OLR} shows that when
$\Omegap=56\Gyr^{-1}$ the region occupied by trapped orbits has moved to
lower $V$ than in the middle panel for $\Omegap=50\Gyr^{-1}$, and is now less
biased to positive $U$. The tail formed by stars that visit four times has
vanished.

\begin{figure}
\centerline{\includegraphics[width=.8\hsize]{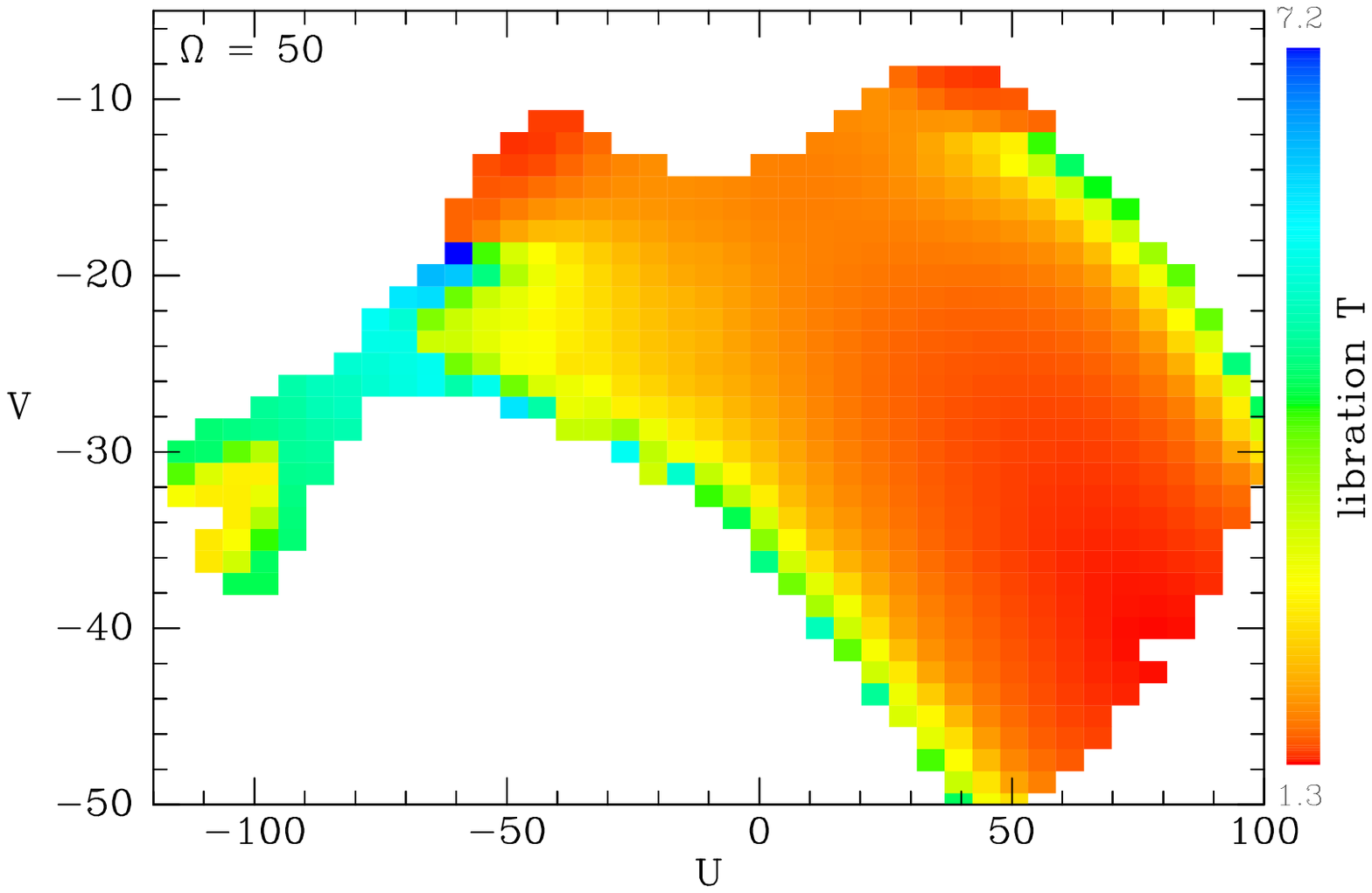}}
\centerline{\includegraphics[width=.8\hsize]{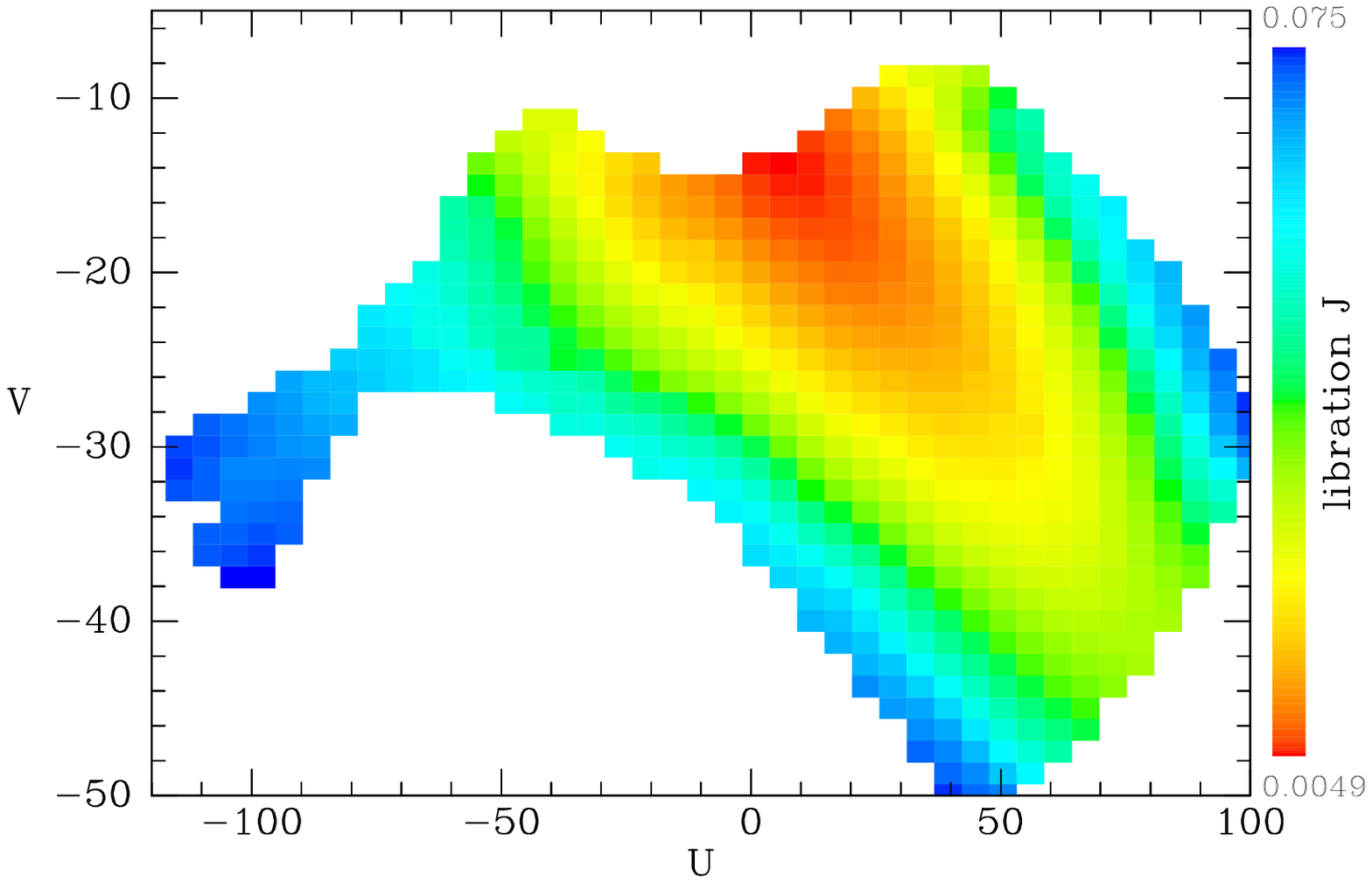}}
\caption{Upper panel: The distribution of libration periods of orbits trapped at OLR when
$\Omegap=50\Gyr^{-1}$. Lower panel: Distribution of libration actions
$\cJ$ of these orbits.}\label{fig:OLR50TJ}
\end{figure}

Even for large values of $\Omegap$, orbits trapped at OLR all have very long
libration periods: the upper panel of Fig.~\ref{fig:OLR50TJ} shows that
with $\Omegap=50\Gyr^{-1}$ the libration periods start at $1.3\Gyr$; with
$\Omegap=56\Gyr^{-1}$ they start at $0.92\Gyr$. Consequently, there is no
compelling case for Jeans' theorem applying to these orbits.

\begin{figure*}
\centerline{\includegraphics[width=.3\hsize]{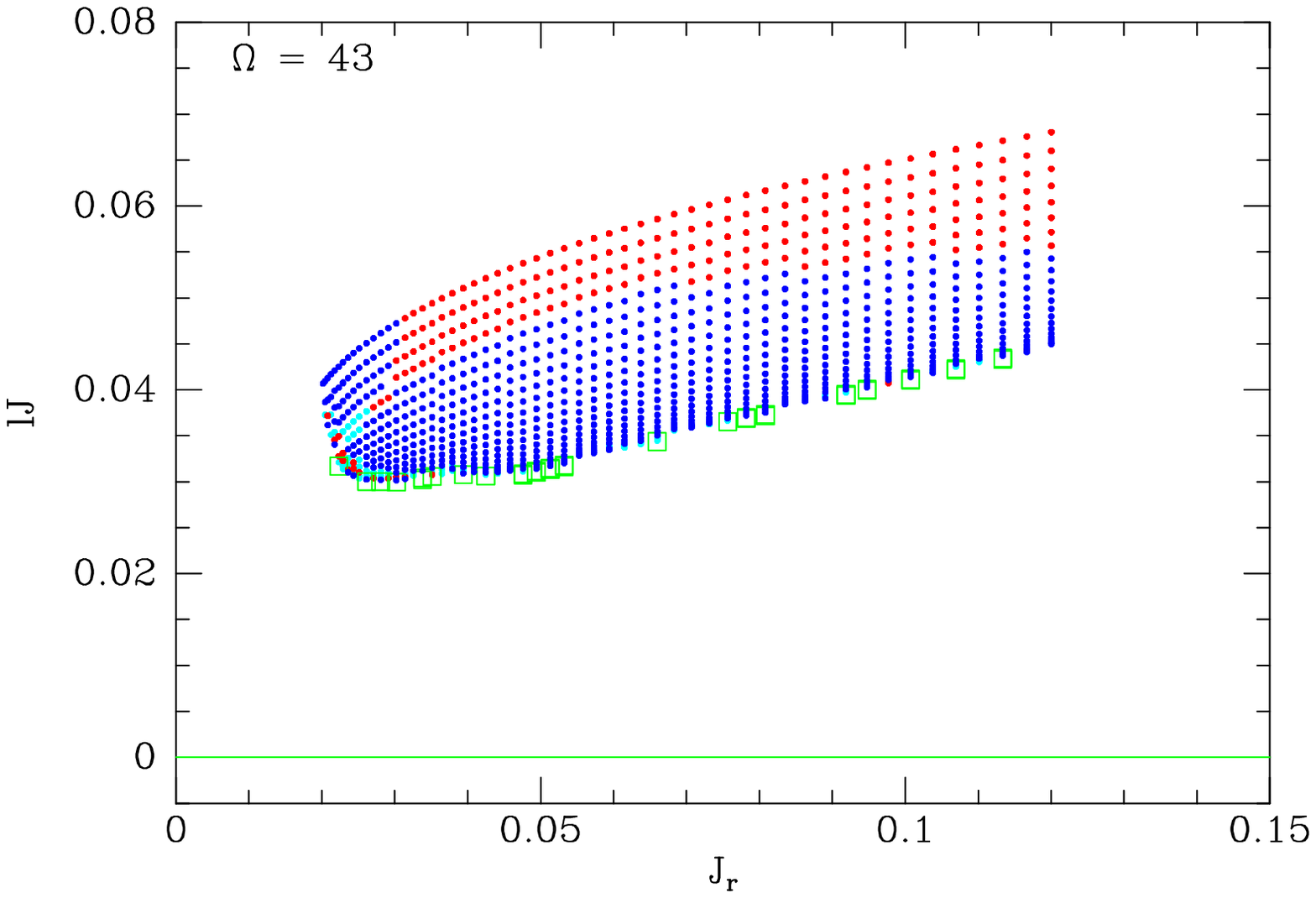}\ 
\includegraphics[width=.3\hsize]{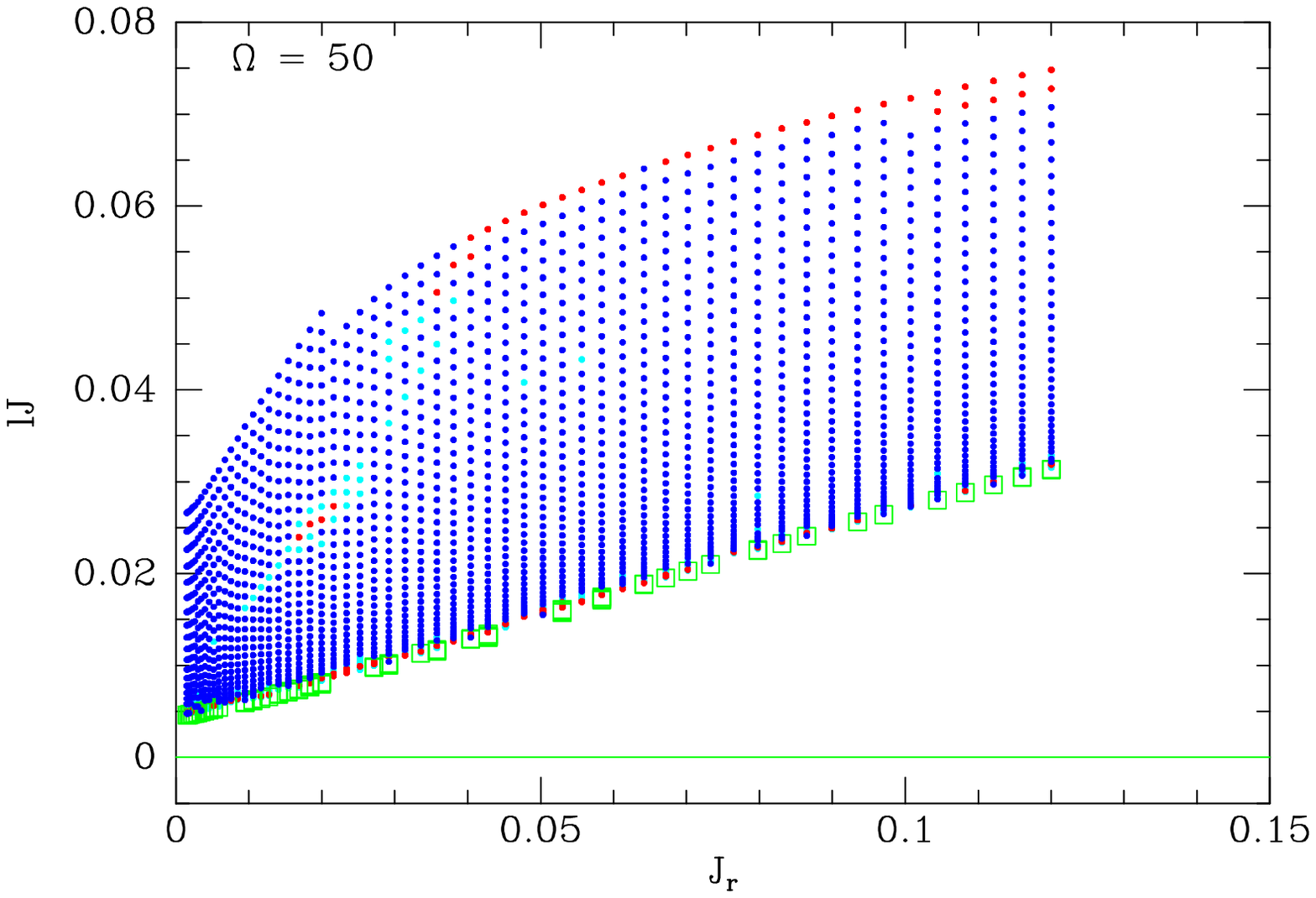}\ 
\includegraphics[width=.3\hsize]{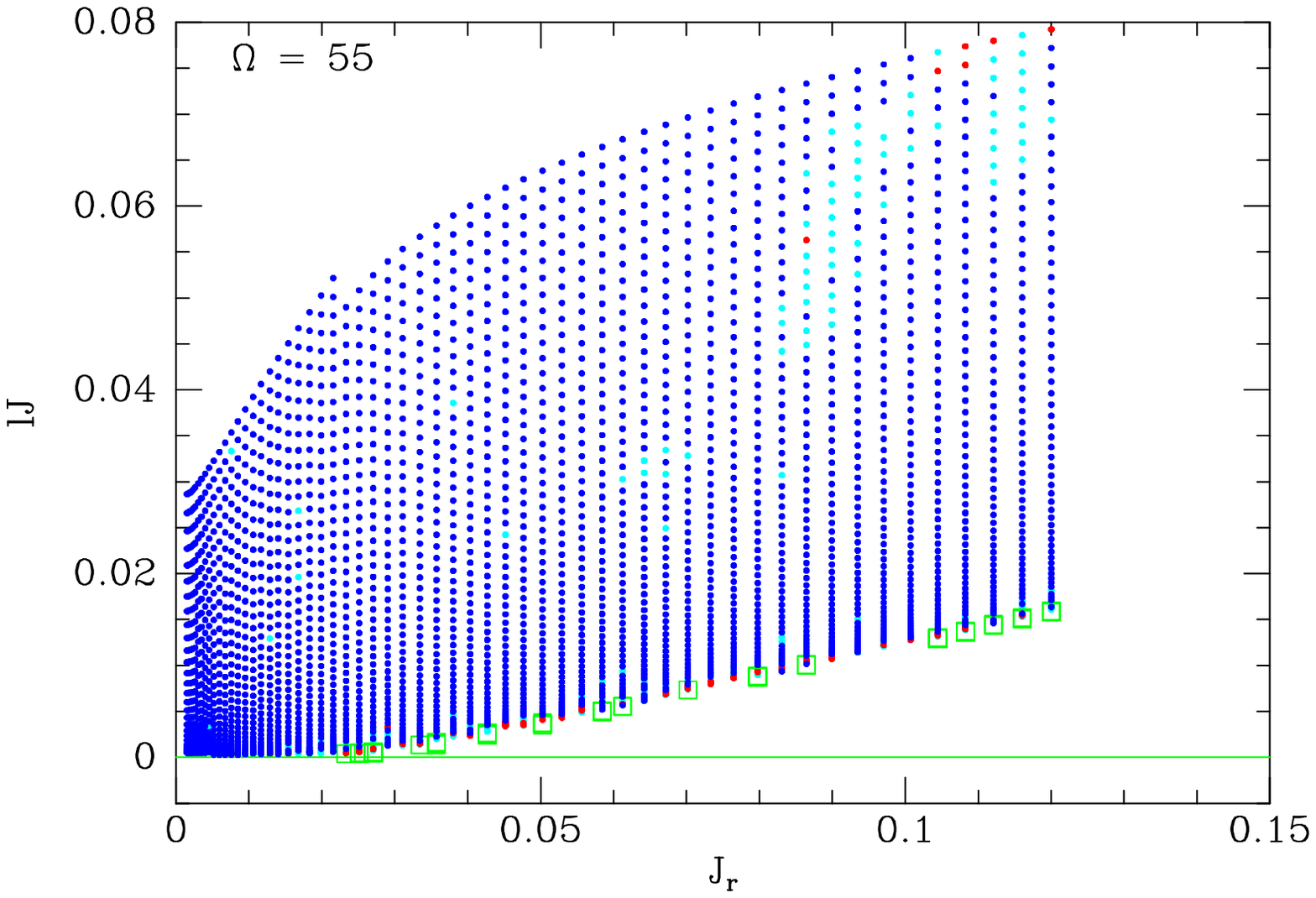}}
\caption{Locations in action space of orbits trapped at OLR that visit the Sun. On the
$x$ axis we have the radial action of the underlying perfectly resonant
orbit: the conserved action $J_3'$ is a function of this
value. Libration actions $\cJ$ are plotted vertically. A blue point indicates
two visits to the Sun with $v_z>0$, a red point indicates four such
visits, and a cyan point indicates one or three visits. A green square indicates a failure to visit.}\label{fig:JJplot}
\end{figure*}

Fig.~\ref{fig:JJplot} shows the location in action space of visiting orbits
for three values of $\Omegap$. As the pattern speed increases, the OLR moves
in towards the Sun. When the OLR is distant, the Sun can be reached only from
eccentric resonant orbits, and then only with a large libration amplitude. As
the OLR approaches, smaller libration amplitudes are required to reach the
Sun, and it becomes easier to reach the Sun from the least eccentric resonant
orbit. Consequently, the block of orbits that visit us moves down and to the
left in Fig.~\ref{fig:JJplot}. In the right two panels, the spikes at
$J_r\simeq0.02$ along the top edges of the populated region signal the
value of $J_r$ at which a clear division between trapped and untrapped orbits
vanishes, as explained in the companion paper.

In Fig.~\ref{fig:DR2OLR} the only intriguing coincidence between a feature in
the observed star density and the structure of a trapped zone is the match in
the middle panel between the most pronounced ridge in the star density and
the lower edge of the zone. However, the statistic $\Delta$ defined by
equation (\ref{eq:defDelta}) indicates that the broader structure of the zone
is incompatible with the hypothesis that the DF is a functions of the
resonant actions: $\Delta=0.51$ when orbits with libration periods
$T_\ell<5\Gyr$ are included and $0.50$ when only orbits with $T_\ell<3\Gyr$
are included. For comparison, the top and bottom panels yield
$\Delta=0.64,\,0.37$ for $T<5\Gyr$ and $0.48,\,0.37$ for $T<3\Gyr$.  Not only
are all these values significantly larger than the values plotted in
Fig.~\ref{fig:sets} for the case of trapping at corotation, but inspection of
the middle panel of Fig.~\ref{fig:DR2OLR} makes clear why $\Delta$ is so
large in the case $\Omegap=50\Gyr^{-1}$: intersections of contours in the
heavily populated, deep blue region $(-30,-18)\kms$ are paired with
intersections of the same contours in the much more sparsely populated light
blue region around $(58,-16)\kms$.

\section{Velocity spaces a kpc away}\label{sec:away}
\begin{figure*}
\centerline{\includegraphics[width=.43\hsize]{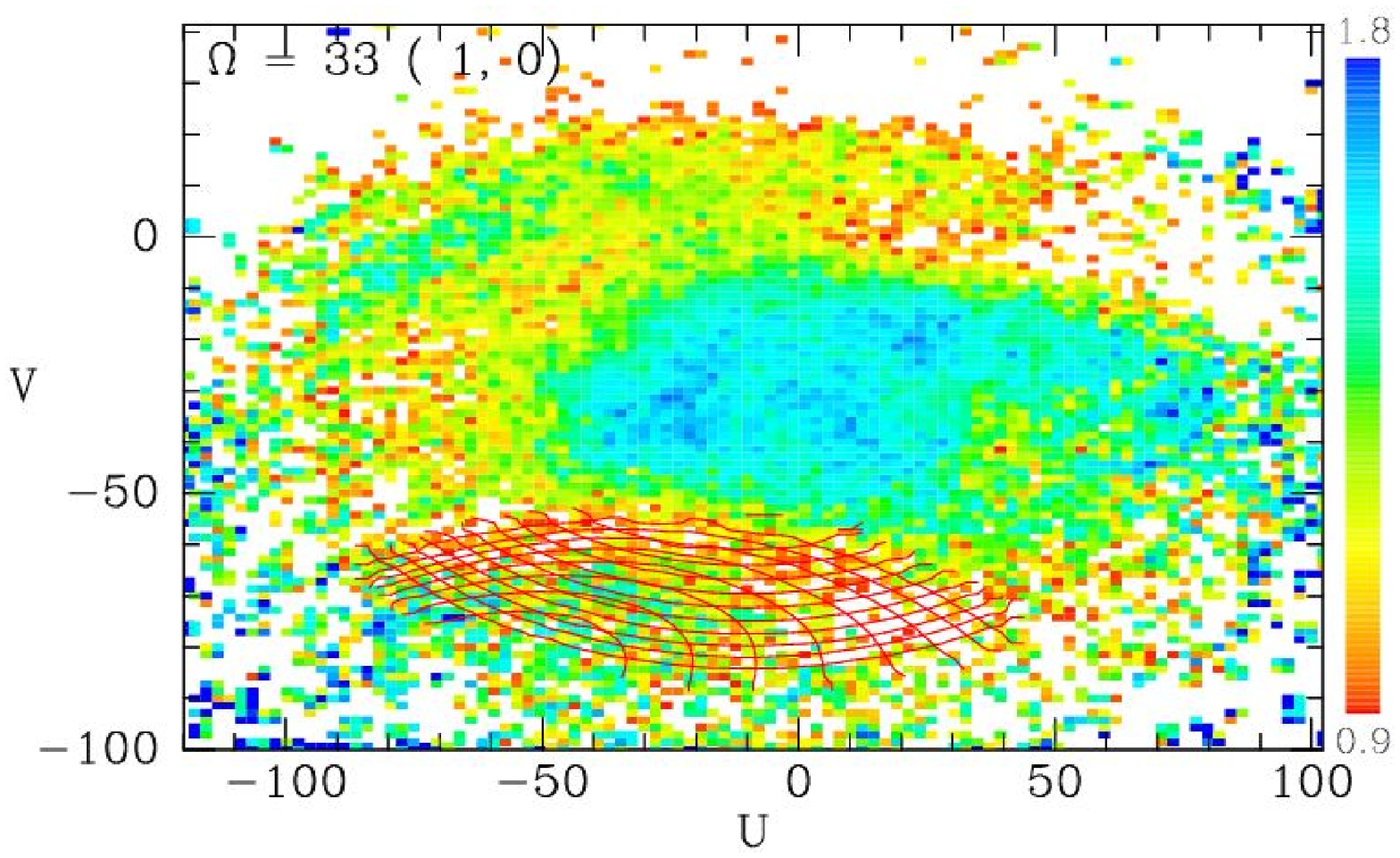}}
\centerline{
\includegraphics[width=.43\hsize]{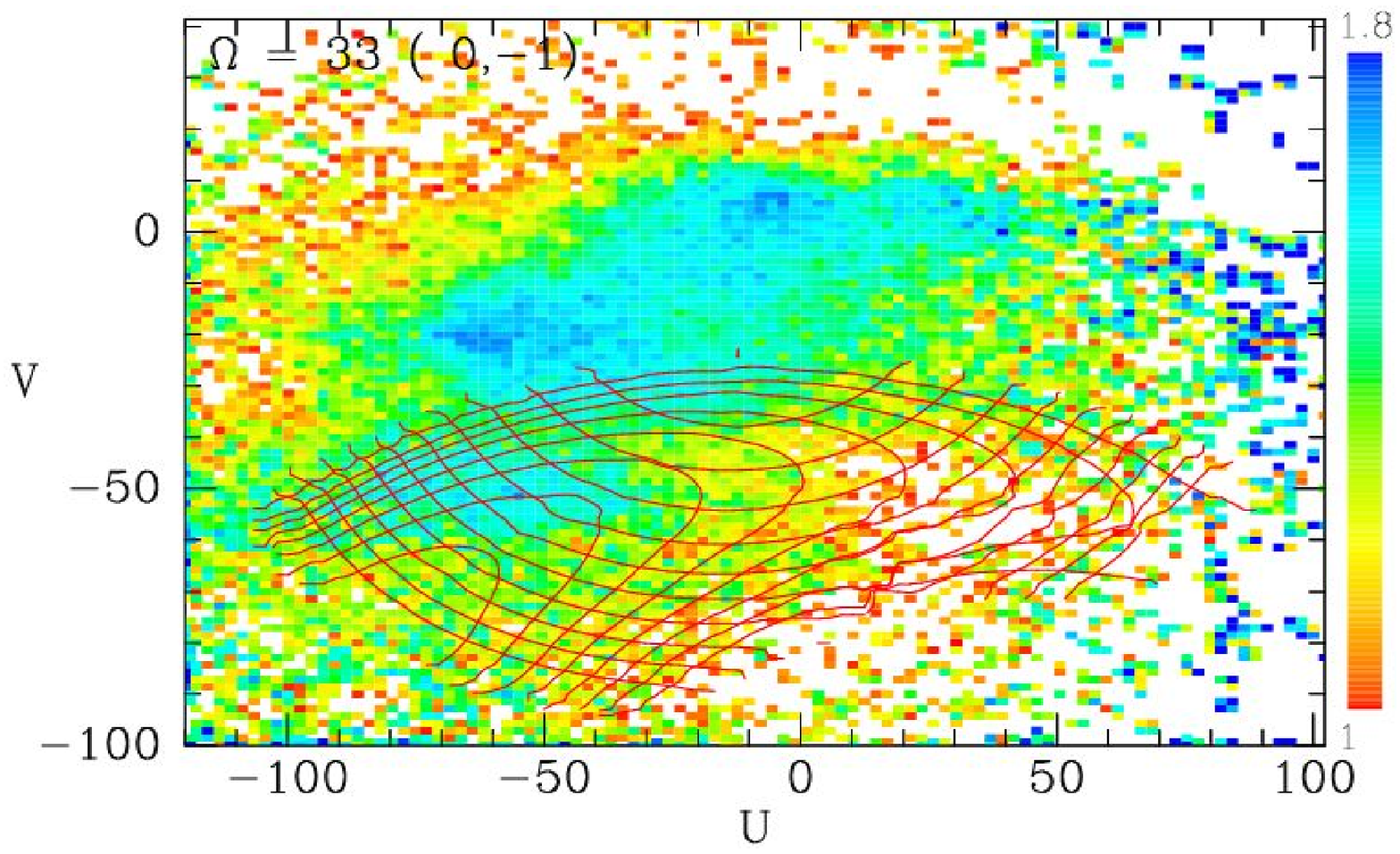}
\includegraphics[width=.43\hsize]{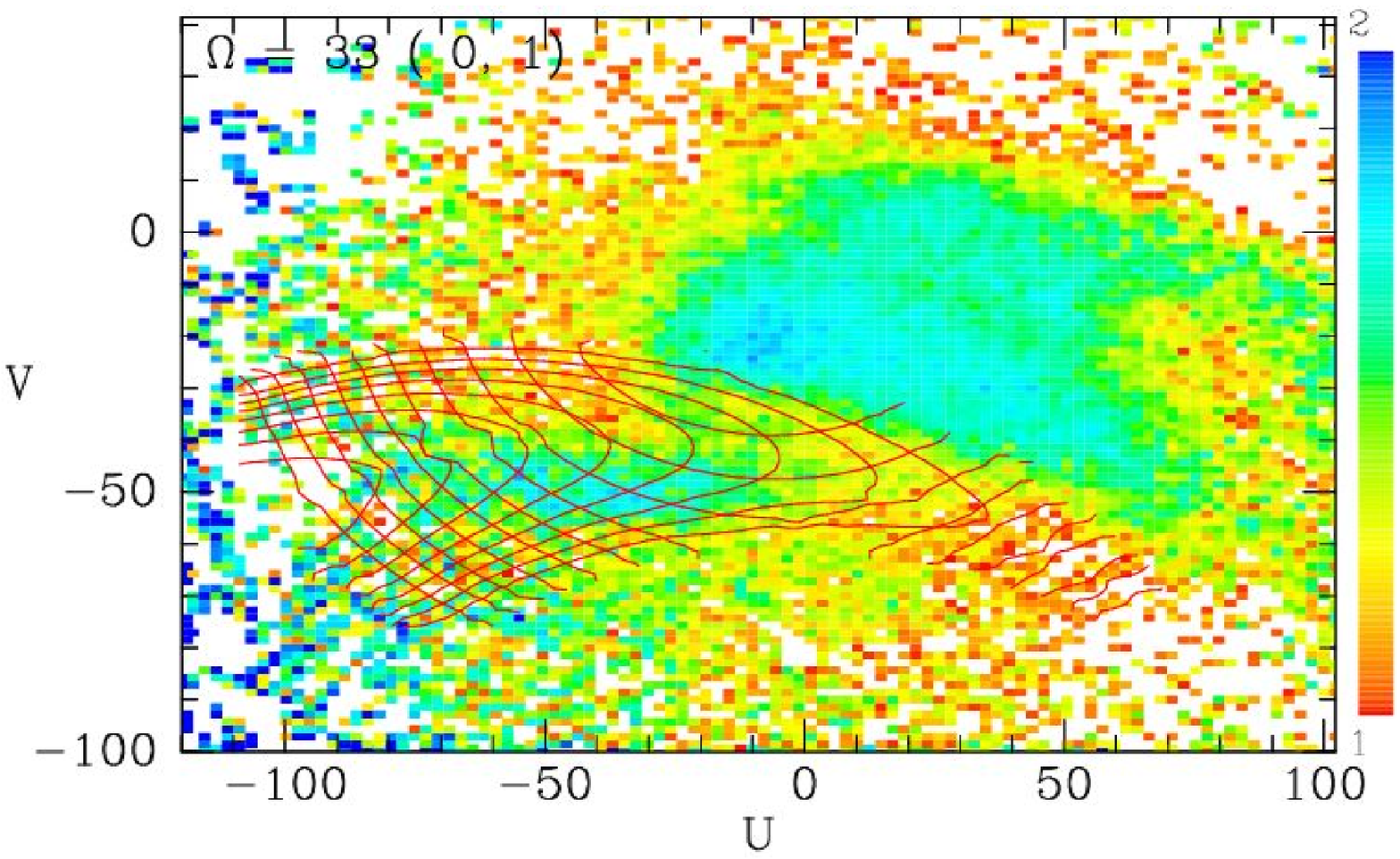}
}
\centerline{\includegraphics[width=.43\hsize]{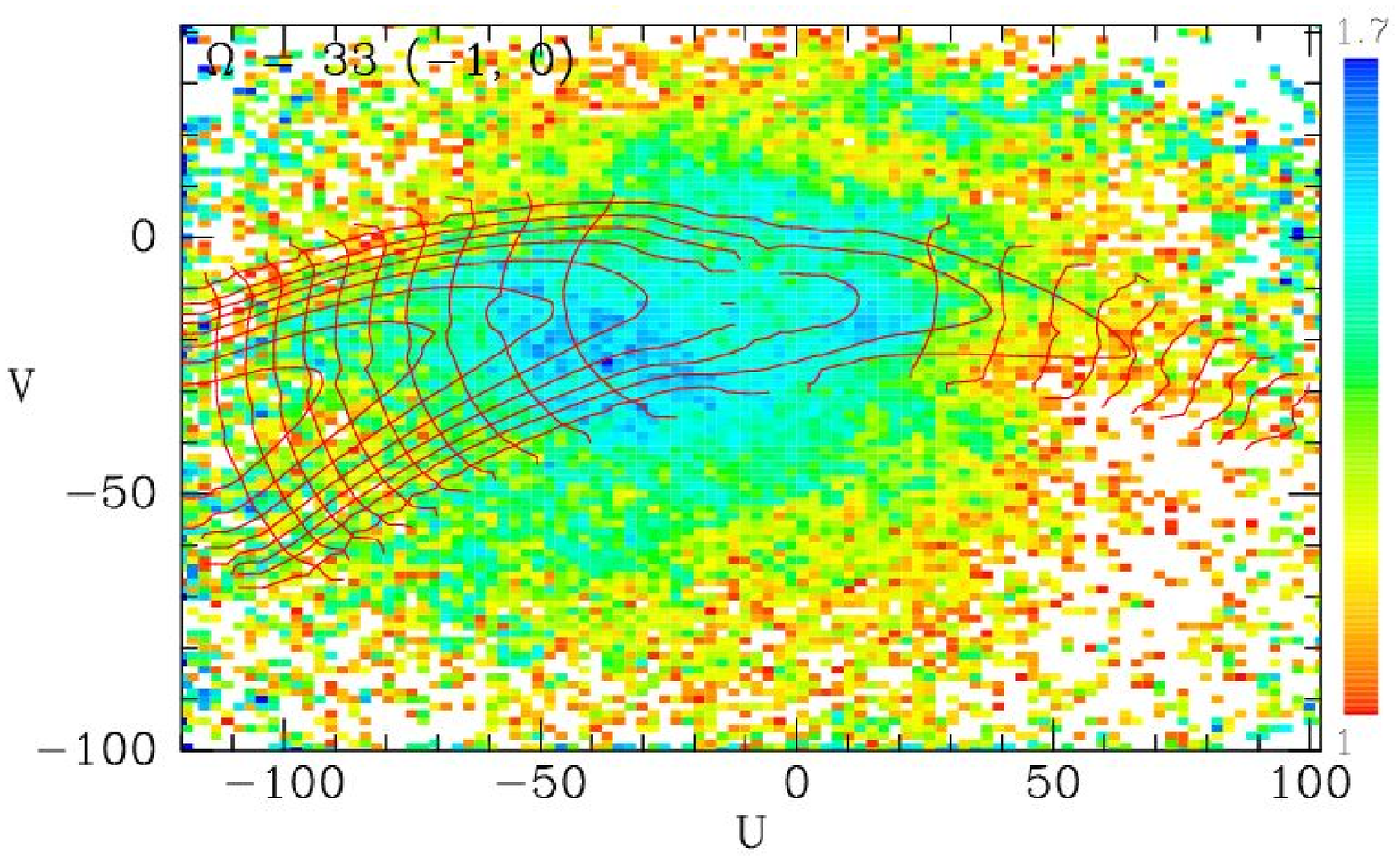}}
\caption{Contours of constant actions when $\Omegap=33\Gyr^{-1}$ superposed
on the density of RVS stars in the velocity space at four locations $1\kpc$
from the Sun. The anticentre lies
upwards and rotation takes stars from left to right.}\label{fig:away}
\end{figure*}

Gaia, unlike Hipparcos, allows us to examine velocity space at locations that
are significantly removed from the Sun. Fig.~\ref{fig:away} shows the star
densities at the four locations reached by moving $1\kpc$
towards or away from the Galactic centre and in or against the direction of
Galactic rotation. In each case stars are included if they lie within
$0.3\kpc$ of the point in question. The contours overplotted are those of
constant actions for trapping at corotation with $\Omegap=33\Gyr^{-1}$. In
these panels the coordinate system is strictly Cartesian, so $U$ is the
component of velocity that is towards the Galactic centre at the Sun, but not
at the locations $(0,\pm1)$ plotted to left and right of the centre panel.
The distribution of stars is centred on negative $U$ at $(x,y)=(0,-1)$ and on
positive $U$ at $(0,1)$ because at these locations the circular velocity has
a non-vanishing $U$ component with that sign. 

Fig.~\ref{fig:away} shows that zones of entrapment move around velocity space
quite rapidly as we move around the Sun. While no striking agreement between
star densities and zones of entrapment is evident, it is in every panel
possible to imagine qualitative similarities. The statistic $\Delta$
defined by equation (\ref{eq:defDelta}) is negative ($-0.38,-0.33$ and
$-0.085$) for three locations
($x=\pm1$ and $y=1$), indicating that deviations from the predictions of
Jeans' theorem are smaller than are expected from Poisson noise. 
By contrast, $\Delta=0.24$ at $y=-1$. This positive value may be no more than
a random upwards fluctuation, but it may reflect a tendency
for velocities $(U,V)\simeq(-20,-65)\kms$  with low densities  to be matched
with $(-70,-45)$ where the density is much higher.

\section{Conclusions}\label{sec:conclude}

An improved version of the torus-mapping code \TM\ has been used to determine
the velocity-space locations of stars that are trapped at the corotation and
outer Lindblad resonances of bars of various pattern speeds. \TM\ computes
the actions of trapped stars so one can add to the observed density of stars
in velocity space contours of the two functions on which the Galaxy's DF
would depend if Jeans' theorem applied to trapped orbits. From intersections
of contours one can identify sets of two, three or four velocities at
which the density of stars would be equal if the adopted potential were
correct and Jeans' theorem applied. Equation (\ref{eq:defDelta}) defines a
measure $\Delta$ of the extent to which the observed star density violates  this
prediction.  A plot of $\Delta$ versus pattern speed weakly supports trapping
at corotation in a bar with pattern speed $(34\pm2)\Gyr^{-1}$. The values
taken by $\Delta$ at locations displaced from the Sun by a kpc in the
cardinal directions are consistent with the hypothesis of trapping at
corotation with $\Omegap=33\Gyr^{-1}$.

The above chain of argument is open to the objection  that the libration
periods of trapped orbits may be too long for Jeans' theorem to apply because
stars have not had time to phase mix. At corotation
libration periods tend to be $\ga1.6\Gyr$ and larger than the radial period
by a factor $\sim12$. An indication that this issue is a real one is the
tendency of $\Delta$ for increase as tori with longer libration periods are
included.

The $\Delta$ statistic offers no support for the hypothesis that trapping at
the OLR is important. A fundamental problem with a connection between the
Hercules stream and trapping at the OLR is that stars trapped in a fast bar
tend to visit the Sun at $U>0$ (approaching the Galactic centre) whereas the
Hercules stream lies at $U<0$. The libration periods of stars trapped at the
OLR of a fast bar are only slightly shorter than the libration periods of
stars trapped at corotation when $\Omegap\simeq33\Gyr^{-1}$.

The impact of resonant trapping on Gaia data has two distinct aspects. One is
the nature of trapped orbits, which has been addressed here. Another is how
those orbits are populated, which is a much harder question.
From Dehnen's seminal (1999) study
onwards most analyses of velocity space have implicitly assumed that the
density of stars can be predicted by assuming adiabatic growth of the bar's
strength at a constant pattern speed. There is no physical basis for this
assumption and the question of how trapped orbits are populated cannot be
answered so simply.

The bar is unlikely to have a constant pattern speed because the latter is a
measure of the angular momentum in the bar, which has both sources and sinks.
Dynamical friction of the bar against the dark halo and the outer disc
removes angular momentum, making the bar slower and longer. Conversely,
angular momentum is stripped from gas that falls through corotaton and is
then funnelled to the central molecular zone \citep{DeSe00,Lia02,Lia03}.
N-body simulations suggest that the bar gradually slows, so its resonances
move out through the disc. 

A star whose orbit lies in the path in action space of a resonance has a
probability to become trapped (and hence transferred to the action space of
trapped orbits), and a probability to be dumped on the far side of the
resonance. The magnitudes of these probabilities depend on the speed at which
the resonance is moving and growing/shrinking. Indeed, if the resonance is
shrinking, stars will always be dumped, while if it is growing strongly while
moving slowly, a star is much more likely to be trapped than dumped. These
considerations imply that the DF within a trapped zone will depend on the
complete history of the relevant resonance and thus of the bar.  

Some stars will have become trapped during bar formation,
which happens essentially on a dynamical timescale \citep{Raha1991}. Such
stars may have been transported a significant distance outwards. Other stars
will have been swept up later as the bar slowed and strengthened. Hence the
density of stars in and around a bar's trapped zone in velocity space must
contain a wealth of information about the history of our disc. 

The natural approach is to follow the dynamics of stars using the
angle-action coordinates of the instantaneous bar. \cite{Chiba2020} have
recently made a start on this enterprise in the context of a simple
razor-thin disc. The task is challenging because both the instantaneous bar
structure and its history have to be inferred.  The technology displayed here
could be used to develop this line of attack. In this context the ability of
\TM\ to model vertical motions in parallel with motion in the plane is likely
to prove important because $J_z$ is known to be strongly correlated with
chemistry and age \citep[e.g.][]{JBH_Galah2019}.

\cite{Monari2019} have recently presented evidence that some of the ridges in the
Gaia DR2 data evident in Figs.~\ref{fig:Gaia33corot} and \ref{fig:DR2OLR} are
associated with resonances driven by $m=3$ and $m=4$ components of the bar's
potential. It would be straightforward to apply \TM\ to these
resonances.

\section*{Acknowledgements}

This work has been supported
by the Leverhulme Trust and the UK Science and Technology Facilities Council
under grant number ST/N000919/1.

\bibliographystyle{mn2e} \bibliography{/u/tex/papers/mcmillan/torus/new_refs}

\end{document}